\documentclass[12pt]{iopart}

\usepackage[english]{babel}
\usepackage[utf8x]{inputenc}
\usepackage[T1]{fontenc}

\usepackage{amssymb,latexsym}
\usepackage{graphicx}
\usepackage[colorlinks=true, allcolors=blue]{hyperref}

\begin{document}

\title{Environmental Noise in Advanced LIGO Detectors}
\author{%
P~Nguyen$^{1}$,  
R~M~S~Schofield$^{1}$,  
A~Effler$^{2}$,  
C~Austin$^{3}$,  
V~Adya$^{4}$,  
M~Ball$^{1}$,  
S~Banagiri$^{5}$,  
K~Banowetz$^{6}$,  %
C~Billman$^{7}$,  %
C~D~Blair$^{8,2}$
% ...
\\
}%
\par\medskip
\address {$^{1}$University of Oregon, Eugene, OR 97403, USA }
\address {$^{2}$LIGO Livingston Observatory, Livingston, LA 70754, USA }
\address {$^{3}$Louisiana State University, Baton Rouge, LA 70803, USA }
\address {$^{4}$Max Planck Institute for Gravitational Physics (Albert Einstein Institute), D-30167 Hannover, Germany }
\address {$^{5}$University of Minnesota, Minneapolis, MN 55455, USA }
\address {$^{6}$LIGO, California Institute of Technology, Pasadena, CA 91125, USA }
\address {$^{7}$University of Florida, Gainesville, FL 32611, USA }
\address {$^{8}$OzGrav, University of Western Australia, Crawley, Western Australia 6009, Australia }
% ...

\date{\today}

\begin{abstract}
    The sensitivity of the Advanced LIGO detectors to gravitational waves can be affected by environmental disturbances external to the detectors themselves. Since the transition from the former initial LIGO phase, many improvements have been made to the equipment and techniques used to investigate these environmental effects. These methods have aided in tracking down and mitigating noise sources throughout the first three observing runs of the advanced detector era, keeping the ambient contribution of environmental noise below the background noise levels of the detectors. In this paper we describe the methods used and how they have led to the mitigation of noise sources, the role that environmental monitoring has played in the validation of gravitational wave events, and plans for future observing runs.
\end{abstract}

\section{Introduction}\label{sec:intro}

Between 2010 and 2015, the two LIGO detectors at Hanford, WA (LIGO Hanford Observatory, or LHO) and Livingston, LA (LIGO Livingston Observatory, or LLO) underwent a period of extensive upgrades to transition from the Initial LIGO stage to the Advanced LIGO (aLIGO) configuration~\cite{aLIGO_design}, significantly improving their sensitivity to gravitational waves~\cite{Martynov_2016}. The aLIGO detectors began their first observing run (O1) on September 12, 2015, and made the first detection of gravitational waves from a binary black hole (BBH) merger on September 14, 2015~\cite{gw150914}, followed by two more BBH detections before the end of the run on January 16, 2016~\cite{gwtc1}. The second observing run (O2) began on November 30, 2016 after a period of detector upgrades and ended on August 25, 2017. During O2, in addition to several more BBH detections, LIGO observed the first binary neutron star (BNS) merger on August 17, 2017~\cite{gw170817}. The third observing run (O3), which spanned April 1, 2019 to March 27, 2020, came after another round of major improvements in the performance of the detectors~\cite{Buikema_2020} and the full inclusion of the Virgo detector in the GW network. In the first half of the run, ending on October 1, 2020, LIGO and Virgo observed a total of 39 GW events~\cite{gwtc2}.

Environmental disturbances can significantly impact the data quality of the LIGO detectors. A gravitational wave (GW) is detected by measuring the differential arm length (DARM) of an interferometer (and converting it to a GW strain), so coupling between the external environment and the interferometer readout can reduce a detector's sensitivity to gravitational waves and potentially produce transient non-astrophysical signals in the detector. The environment can influence the detector through physical contact (via vibrations or temperature fluctuations), electromagnetic waves, static electric and magnetic fields, and possibly high-energy radiation. These effects are monitored with the physical environmental monitoring (PEM) system of sensors~\cite{Effler_2015}.

Studying environmental noise serves two purposes. The first is the validation of GW events. Environmental disturbances at amplitudes large enough to influence the LIGO data occur frequently around each detector and can potentially be correlated between different detectors, i.e. stemming from a common source as opposed to stemming from chance coincidence. Such correlated noise is not accounted for in the estimation of false-alarm probabilities, which is done by time-shifting background data from each LIGO detector to produce long stretches of coincident background. Environmental noise is particularly important in searches for un-modeled sources of gravitational waves, as these look for excess power without the use of waveform templates. Thus it is important to have a quantitative solution for identifying and evaluating the impact of environmental transients when they occur near candidate events.

The second purpose is to improve the sensitivity of the detector by reducing contamination from environmental noise. We track down troublesome noise sources and coupling mechanisms so that we can either remove the noise sources themselves, isolate them from the detector, or modify the detector to reduce coupling.

Effler et al.\ (2015)~\cite{Effler_2015} described the methodology for studying environmental coupling and presented results from the sixth and final science run (S6) prior to the transition to aLIGO. The methodology has since been improved and expanded, and sensitivity to environmental effects has changed with the upgrades to the detector. This paper describes these changes and presents cases where noise sources have been identified and mitigated between S6 and the end of O3. We also summarize how GW events are vetted using quantitative results from injections. This paper focuses on noise investigations at the LIGO detectors; a similar discussion for the Virgo detectors is provided in Fiori et al.\ (2020)~\cite{Fiori_2020}.

There are many techniques for characterizing detector noise beyond those described here~\cite{Davis_2021, Nuttall_2015, Nuttall_2018, Detchar_2016, Davis_2019}. These include the use of tools for detecting excess power transients in the strain data~\cite{Chatterji_2004}, categorizing transients using machine learning to better distinguish them from astrophysical signals~\cite{Zevin_2017, Bahaadini_2018}, searching for correlated noise between auxiliary sensors and the strain data~\cite{Smith_2011}, and many more. Although these also play a role in achieving the goals above, this paper discusses more direct, focused techniques for studying, quantifying, and mitigating environmental effects.

This paper is organized as follows. In Section~\ref{sec:upgrades} we summarize the changes made to the LIGO detectors and the PEM system since S6. In Section~\ref{sec:analysis} we present a method for quantifying environmental coupling based on data from noise injections. In Section~\ref{sec:injections} we describe developments in the techniques for performing environmental noise injections. In Section~\ref{sec:O3} we show results of recent studies and provide examples of how environmental influences have been mitigated. In Section~\ref{sec:vetting} we describe the process of vetting GW event candidates with examples from real events. We conclude with a discussion of future work in Section~\ref{sec:conclusion}.

\section{aLIGO Upgrades}\label{sec:upgrades}

\subsection{Detector Upgrades}\label{sec:detector_upgrades}

\begin{figure}
    \centering
    \includegraphics[width=0.8\textwidth]{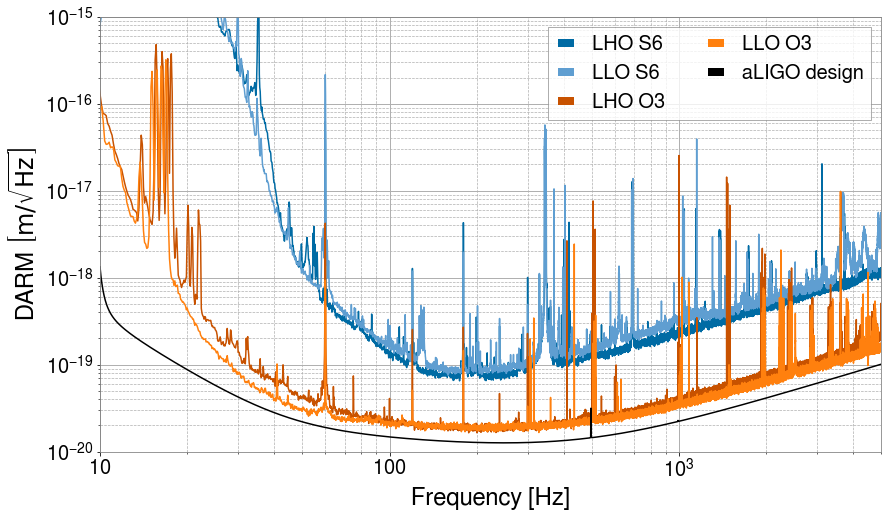}
    \caption{Amplitude spectral densities of the differential arm length displacement (DARM) at the end of S6 (Feb 27 2010 04:27:47 UTC) and during O3 (Mar 20 2020 00:00:00 UTC).}
    \label{fig:darm}
\end{figure}

When fully commissioned aLIGO is designed to provide an order of magnitude improvement to sensitivity in its most sensitive band~\cite{aLIGO_design}, as well as more than an order of magnitude improvement at lower frequencies due to seismic isolation upgrades. \Fref{fig:darm} compares the DARM noise spectra of LHO and LLO at the end of S6 to that at the end of O3. Significant progress has been made in approaching the design sensitivity of aLIGO, and further improvements are foreseen for the fourth observing run (O4), expected to begin in 2022. Here we highlight a few of the major upgrades that were directly relevant to reducing the coupling of ambient environmental noise.

The core interferometer optics (including the test mass mirrors and beam splitter) are suspended in active multi-stage suspension systems, which in turn are on active seismic isolation tables~\cite{Robertson_2002, Aston_2012}. This provided a substantial improvement to sensitivity below 100\,Hz over the initial LIGO configuration. The suspension and isolation tables also provide useful sensors for the motion of these optics.

Auxiliary sensors used in the control system of the interferometer were moved from in-air optical tables to in-vacuum, seismically isolated tables. This reduced acoustic coupling but did not eliminate it.  Although the main laser system (PSL, or pre-stabilized laser) could not be moved into vacuum, an acoustically isolated room was built to house the laser, and a new optical table with improved isolation and damping of its resonances was installed~\cite{Schofield_2011}.

To reduce magnetic coupling, magnets and certain magnetic materials are no longer present on or near the test masses themselves~\cite{Schofield_2010}. Instead, the aLIGO test masses are controlled either by magnets at the upper stages of the suspension system or by an electrostatic drive. To further reduce the ambient acoustic noise from electronics fans near the detector, power supplies and most electronics were moved to separate rooms (called here electronics bays), some tens of meters away from the vacuum system which houses the interferometer.

\subsection{Environmental Monitoring Upgrades}\label{sec:pem_sensors}

Understanding environmental influences on the detectors requires comprehensive monitoring of its physical surroundings. This is done through the PEM system of auxiliary sensors, which consists of accelerometers for high-frequency vibrations (tens to thousands of Hz), seismometers for low-frequency vibrations (up to tens of Hz), microphones, magnetometers, voltage monitors that measure the voltage of electric power supplied to the detector sites, radio-frequency (RF) receivers, a cosmic-ray detector for high-energy particles, and wind, temperature and humidity sensors. Detailed information on PEM sensors, including example background spectra and calibration data, can be found on the PEM website, PEM.LIGO.org~\cite{PEM_website}. The site also provides links to long-term summaries of ground tilt, seismic motion, and wind (on the Environmental Studies pages).

In order to monitor environmental signals that could influence the interferometer, we use PEM sensors that are demonstrated to be much more sensitive to these signals than the interferometer is. Sensor locations are chosen with the goal of maximizing coverage of potential coupling sites. Ideally, if an environmental signal were to reach a coupling site, nearby sensors should be able to observe the signal at an amplitude equal to the amplitude at the coupling site. In practice, we place sensors where we expect the coupling to be strongest, and we may place new sensors during the run to improve monitoring of important coupling sites.

By focusing on the fundamental interactions that can affect the detector, the PEM system allows us to monitor potential effects from a large variety of environmental events. For example, wind can couple through vibrations in the ground and air, so its effects are monitored by seismometers, accelerometers, and microphones. Lightning could couple by magnetic fields, power mains disturbances, and electromagnetic waves at radio frequencies that we demodulate into the detection band, so lightning strikes are monitored with magnetometers, mains monitors, and RF receivers. The PEM sensors provide coverage of signals in the detection band of the interferometer (20-2000\,Hz), although we also monitor beyond these frequencies when there are coupling mechanisms that convert low- or high-frequency signals up or down into the detection band or when the interferometer performance can be affected by frequencies outside of the detection band.

\begin{figure}
    \centering
    \includegraphics[width=\textwidth]{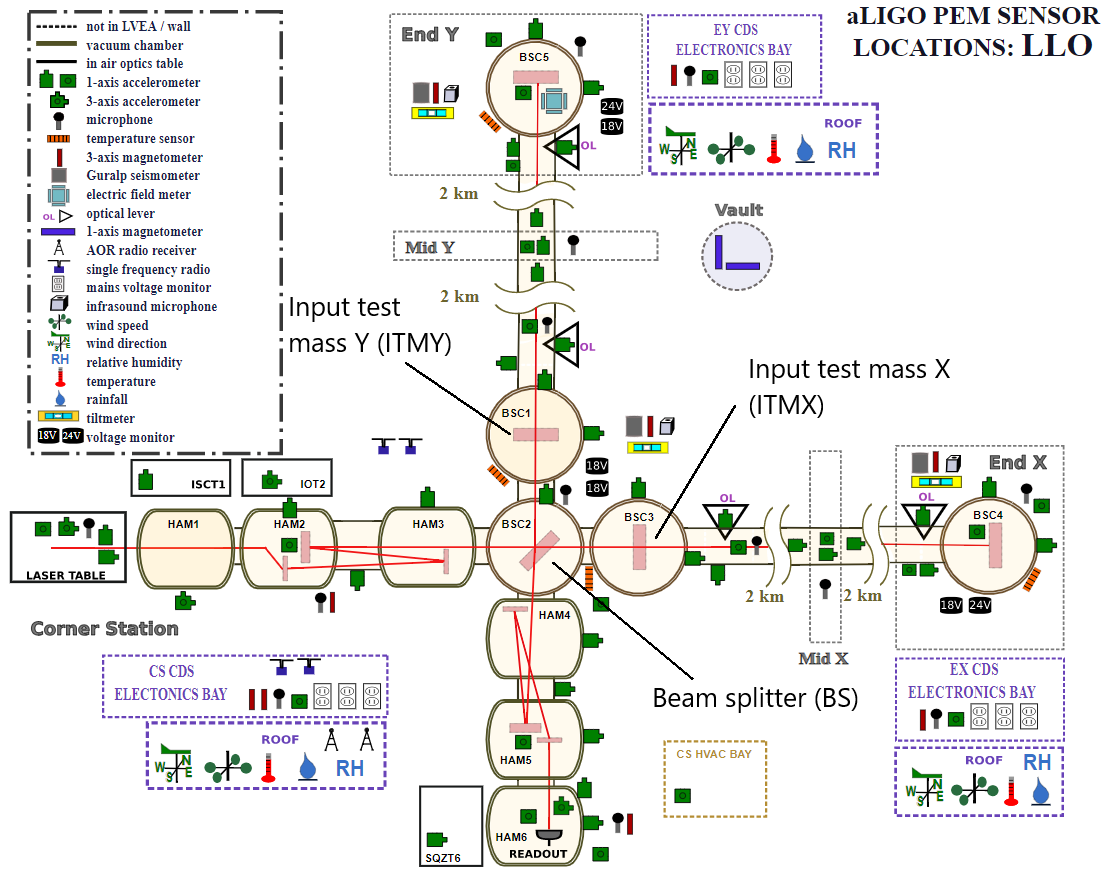}
    \caption{The Physical Environmental Monitoring system layout at the LIGO Livingston detector during O3, as seen on the PEM public website~\cite{PEM_website}. The path of the main interferometer laser is shown as a red line; core optics, such as the test masses, are represented by rectangles inside the vacuum chambers. The most major changes during aLIGO have been made to the accelerometer locations and the addition of new magnetometers, e.g. in the electronics bays.}
    \label{fig:sensors}
\end{figure}

The state of the PEM system at LLO during O3 is shown in \Fref{fig:sensors}. A similar map for LHO is available on the PEM website~\cite{PEM_website}. Since the transition to aLIGO, many changes~\cite{Schofield_2012} have been implemented to expand the general coverage of the PEM system, provide additional monitoring near known high-coupling areas, and adapt to the detector upgrades described in Section~\ref{sec:detector_upgrades}. The density of accelerometers and magnetometers at each observatory doubled between the end of S6 and the end of O3. In its O3 configuration, LHO has 62 accelerometers, 14 microphones, and 9 magnetometers; LLO has 64 accelerometers, 13 microphones, and 9 magnetometers.

Many of the changes since initial LIGO involved the addition of accelerometers or relocation of existing ones:

\begin{itemize}
    \item Vacuum chambers: In iLIGO, most accelerometers were mounted on the seismic isolation system. These locations became redundant with the introduction of vibrational sensors as part of the new active isolation systems, so the accelerometers are now mounted on the chamber walls where they can detect motions that could modulate laser light scattered off of the chamber walls.
    \item Beam tubes: Accelerometers at select sites along the 4-km beam tubes monitor vibrations that affect the modulation of reflected light inside. Coverage now includes the mid-stations, which is especially important at LHO where significant coupling has been measured, likely because they contain the smallest aperture between vertex and end stations.
    \item Electronics bays: Floor accelerometers were added to detect vibrational coupling to the electronics boards (e.g. through resistance variations in poor solder joints) and to monitor the rooms as seismic sources.
    \item Vacuum enclosure areas: Floor accelerometers were added near the vacuum chambers in order to expand coverage and aid in localizing sources of vibrations through propagation delays and amplitude differences at locations that do not have the resonance structure of the vacuum envelope.
    \item Pre-stabilized laser table: Coverage of the main laser table was expanded. This area has continued to be a major source of vibrational coupling.
\end{itemize}

Many sensors were also upgraded to newer models in order to improve their reliability, since many of them were outdated. Table~\ref{tab:sensors} summarizes the current sensor models and specifications.

Magnetometer coverage was also expanded, particularly with the addition of magnetometers on the electronics racks (located in the electronics bays) which were important noise sources and coupling sites during initial LIGO. Relatively large magnetic fields are generated by the equipment in the racks and these fields can couple to components, cables and, connectors in the racks. Additionally, magnetometers in electronics racks have been useful for identifying sources of narrow spectral peaks even when the coupling was not through magnetic fields. Cyclical processes producing line artifacts in the DARM spectrum can be tracked down by detecting currents associated with those processes. In a sense, we monitor multiple electronic systems at once, using fluxgate magnetometers  in the electronics racks (Bartington-03 series~\cite{bartington}). These are sensitive enough to detect periodic currents with amplitudes as low as $5\times10^{-5}$\,A at 1\,m from long wires or traces~\cite{Covas_2018}.

Additionally, the non-rigid tripods for fluxgate magnetometers were replaced with rigid ones. Non-rigid tripods lead to increased cross-talk between floor vibrations and the magnetometer signal, as the magnetometer vibrates relative to the Earth's magnetic field at the tripod resonance frequency. This created a narrow 7.9\,Hz peak in the magnetometer spectrum a factor of three above background, which was eliminated by switching to rigid surveyors' tripods~\cite{alog_tripod}.

In addition to the fluxgate magnetometers that monitor local magnetic fields, extremely low frequency induction coil magnetometers (LEMI-120~\cite{lemi}) were added to the PEM system in order to monitor magnetic noise from Schumann resonances. These are global electromagnetic resonances in the cavity formed by the Earth's surface and the conductive ionosphere. Lightning strikes around the world excite this resonant cavity, producing picoTesla-scale magnetic fields that can cause correlated noise in the LIGO detectors~\cite{Thrane:2013npa, Thrane:2014yza, Coughlin_2018}. Two LEMI magnetometers are positioned at each site, far enough from the detector so that they are not sensitive to the same local magnetic fields observed by the fluxgate magnetometers. They are placed at a location between the corner station and end stations, 100-200\,m from the beam tube, one aligned with the $x$-axis and one with the $y$-axis of the interferometer.

An electric field meter was installed in an end test mass chamber at each observatory~\cite{Abbott_2018, alog_efm, Buikema_2020}. These can detect electric fields generated inside of the chambers as well as fields from outside the chamber that make it in through glass viewports on the chambers.

\begin{table}
    \caption{\label{tab:sensors}Specifications for important PEM sensor types. The operating frequency range is the range in which the sensor calibration is flat; we often use them over a broader range. Noise floor numbers are reported in the operating band of each sensor (seismometer at 1\,Hz).}
    \small
    \begin{tabular}{@{}lllll}
        \hline
        Type & Sensor & Operating freq. & Sampling freq. & Noise floor \\
        \hline
        \hline
        seismometer & Guralp\textsuperscript{\textregistered} CMG-3T~\cite{guralp, Peterson_1993} & 0.1-20\,Hz & 256\,Hz & <1 nm/s$/\sqrt{\textrm{Hz}}$ \\
        \hline
        accelerometer & Wilcoxon\textsuperscript{\textregistered} 731-207~\cite{wilcoxon} & 1-900\,Hz & 4096\,Hz & 0.5 $\mu$m/s$^2/\sqrt{\textrm{Hz}}$ \\
        \hline
        microphone & Br\"{u}el\&Kj\ae r\textsuperscript{\textregistered} 4130~\cite{bk4130} & 10-900\,Hz & 16384\,Hz & <30 $\mu\textrm{Pa}/\sqrt{\textrm{Hz}}$ \\
        \hline
        microphone & Br\"{u}el\&Kj\ae r\textsuperscript{\textregistered} 4188~\cite{bk4188, bk4188_powersupply} & 8-12500\,Hz & 16384\,Hz & <5 $\mu\textrm{Pa}/\sqrt{\textrm{Hz}}$ \\
        \hline
        magnetometer & Bartington\textsuperscript{\textregistered} 03CES100~\cite{bartington} & 0-900\,Hz & 8192\,Hz & <6 pT/$\sqrt{\textrm{Hz}}$ \\
        \hline
        magnetometer & LEMI-120\textsuperscript{\textregistered}~\cite{lemi} & 0.0001-1000\,Hz & 4096\,Hz & <0.1 pT/$\sqrt{\textrm{Hz}}$ \\
        \hline
        radio station & AOR\textsuperscript{\textregistered} AR5000A~\cite{aor} & 24.5 MHz & 16384\,Hz & ---\\
        \hline
    \end{tabular}
\end{table}

\section{Coupling Functions}\label{sec:analysis}

To determine the degree to which the detector is affected by environmental influences during operation, we inject basic environmental disturbances that produce a response in DARM. We make acoustic injections with speakers and monitor them with the system accelerometers and microphones; seismic injections with shakers, monitoring them with the accelerometers and seismometers; magnetic injections with wire coils monitored with the magnetometers. The injection methodology is described in more detail in Section~\ref{sec:injections}. To motivate the injection techniques we first discuss the means of quantifying the coupling.

Suppose there exists only one coupling site, a sensor is placed at the location of the coupling site, and a noise injection is performed that produces a signal in the sensor and some response in DARM. A \textit{coupling function} can be computed based on the actuation measured by the witness sensor and the response measured in DARM~\cite{Kruk_2016, pem_code}. We compare the amplitude spectral densities (ASDs) of DARM and the witness sensor during the time of the injection (\textit{injection time}) to their ASDs during a time when both are at observation-mode noise levels (\textit{background time}). The coupling function at some frequency $f$ is given by

\begin{equation}\label{eq:cf}
    \mathrm{CF}(f) = \sqrt{\frac{[Y_{\textrm{inj}}(f)]^2 - [Y_{\textrm{bkg}}(f)]^2}{[X_{\textrm{inj}}(f)]^2 - [X_{\textrm{bkg}}(f)]^2}}
\end{equation}
where $X_{\textrm{bkg}}(f)$ and $X_{\textrm{inj}}(f)$ are the ASDs of the witness sensor at background and injection times, respectively, and $Y_{\textrm{bkg}}(f)$ and $Y_{\textrm{inj}}(f)$ are the ASDs of DARM at background and injection times. We use \textit{coupling factor} to refer to the value of a coupling function at a single frequency bin.

A sensor's coupling function can be used to compute the contribution of noise in the sensor to DARM. For example, when validating GW events, we multiply the coupling function by the amplitude of any environmental transient observed by the sensor to predict the corresponding amplitude in DARM. Additionally, multiplying the coupling function by the sensor's ambient background level yields the ambient contribution of noise at the sensor to the DARM spectrum: $Y(f) = \mathrm{CF}(f) X(f)$.

Suppose now we expand the scenario such that there are multiple coupling sites, and a sensor is placed at the location of each site. We can model the response in DARM to each injection as a linear combination of the sensor signals and their sensor-specific coupling functions. To solve for the coupling functions, we can perform multiple injections instead of just one, resulting in a system of $n$ equations with $m$ unknown coupling functions, where $n$ and $m$ are the numbers of injections and sensors, respectively:

\begin{equation}\label{eq:cf_full}
    Y_i(f) = \sum_{j=1}^{m} \mathrm{CF}_j(f) X_{ij}(f).
\end{equation}
Here $Y_i(f)$ and $X_{ij}(f)$ are the amplitudes of DARM during injection $i$ and sensor $j$ during injection $i$ respectively, and $\mathrm{CF}_j(f)$ is the coupling function of sensor $j$. One could solve~\eref{eq:cf_full} to determine the coupling functions of all sensors.

We have assumed thus far that the witness sensors are placed at the locations of the coupling mechanisms, but such perfect placement is not realistically feasible given that there are an unknown number of coupling sites at unknown locations. A sensor, even if it is near a coupling site, only measures the injection amplitude at its own location, not at the coupling location. Therefore, when using real-world sensors,~\eref{eq:cf} is only an estimate of the true coupling, and \eref{eq:cf_full} is not an exact model of all the coupling mechanisms. Nevertheless, as explained above, we distribute sensors to maximize coverage of coupling sites and find that this has been sufficient for producing reliable coupling functions for all sensors, as discussed further in Section~\ref{sec:uncertainties}.

One hurdle remains in attempting to solve~\eref{eq:cf_full}. In practice, typically $n<m$ due to logistical constraints on the number of injections one could perform during a realistic time window, which makes the system of equations underdetermined. The problem can be simplified by instead approximating $\mathrm{CF}_j(f)$ for each sensor independently of other sensors. Given a sensor $j$, we can re-purpose \eref{eq:cf} (replacing $X$ with $X_{ij}$ and $Y$ with $Y_i$) to compute a single-injection ``coupling function'' $\mathcal{CF}_{ij}(f)$ for each injection, then combine those to produce an approximation to $\mathrm{CF}_j(f)$. The closer an injection is to a sensor, the more accurate the computed $\mathcal{CF}_{ij}(f)$ would be to $\mathrm{CF}_j(f)$, since the DARM response would be dominated by coupling near sensor $j$. Since it is impractical to produce an injection at each sensor, the approach we have adopted for combining the $\mathcal{CF}_{ij}(f)$ is to construct a \textit{composite coupling function} whose value at each frequency bin is the coupling factor corresponding to the nearest injection, determined by the highest sensor amplitude (using the assumption that injection amplitudes are equivalent). That is, for a frequency $f_k$ and a set of injections $\mathcal{I}$, we measure the sensor amplitudes $\{X_{ij}(f_k)\ |\ i \in \mathcal{I}\}$, compute the single-injection coupling functions $\{\mathcal{CF}_{ij}(f_k)\ |\ i \in \mathcal{I}\}$, and compute the composite coupling function as

\begin{equation}\label{eq:ccf}
    \widetilde{\mathrm{CF}}_j(f_k) := \mathcal{CF}_{lj}(f_k)\ \mathrm{where}\ l = \mathop{argmax}_{i\in\mathcal{I}}\ (X_{ij}(f_k)).
\end{equation}

If the distribution of injection locations provides sufficient coverage of sensor locations, then $\widetilde{\mathrm{CF}}_j(f) \approx \mathrm{CF}_j(f)$. We discuss shortcomings of this assumption in Section~\ref{sec:uncertainties}.

\begin{figure}
    \centering
    \includegraphics[width=\textwidth]{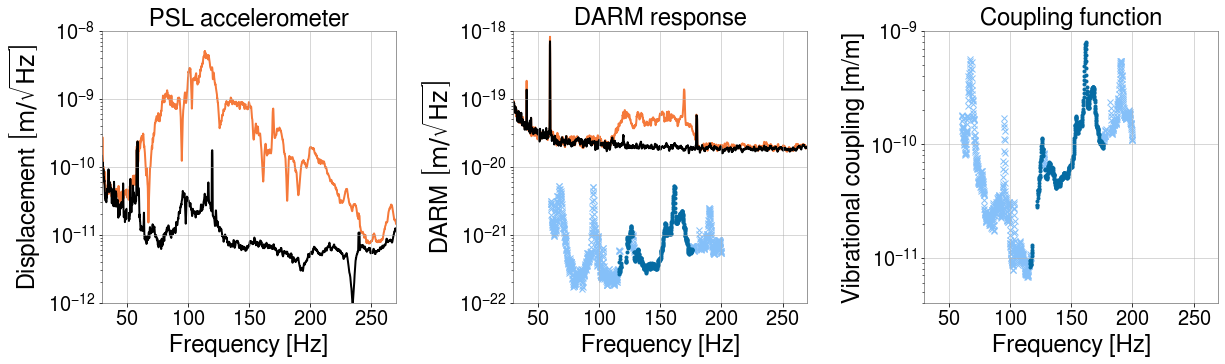}
    \caption{Vibrational coupling excited by a broadband (60-200\,Hz) acoustic injection near the output arm of the interferometer (HAM5 and HAM6 in \Fref{fig:sensors}). The left plot shows the displacement of an accelerometer in the PSL room during background time (black) and injection time (orange). The middle plot shows the interferometer readout during background time (black) and injection time (orange). Estimated ambient levels for the accelerometer are also shown as dark blue dots, with upper limits shown as light blue crosses; they are produced from the single-injection coupling function in the right plot. A vibrational single-injection coupling function represents meters of differential test mass displacement per meter of sensor displacement, hence the units of m/m.}
    \label{fig:injection}
\end{figure}

Computing the single-injection coupling functions $\mathcal{CF}_{ij}(f)$ (example shown in \Fref{fig:injection}) requires a significant difference between the injection and background signals in the sensor and in DARM. To distinguish between measurements and upper limits, thresholds are chosen for the sensor and DARM in the form of a ratio between the injection ASD and background ASD. For each frequency bin, if an injection produces a large enough signal to exceed both the sensor threshold and DARM threshold, then a coupling factor can be measured via Eq.~\ref{eq:cf}. If the injection exceeds the sensor threshold but not the DARM threshold, then we instead compute an upper limit by omitting the DARM background term. These thresholds are typically chosen to be a factor of two in DARM and a factor of a few in the sensor, based on the typical level of fluctuations observed in the spectra.

\begin{figure}
    \centering
    \includegraphics[width=\textwidth]{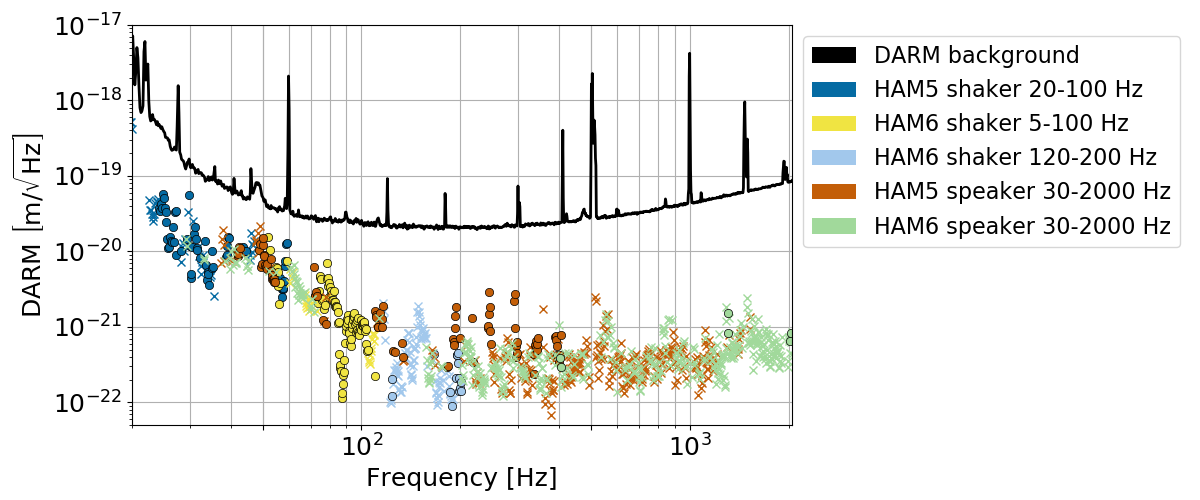}
    \caption{Ambient noise level for the LHO HAM6 Y-axis accelerometer estimated from a composite coupling function, using acoustic and seismic injections near the output arm. For simplicity only five injections were used to produce this example, however in practice the number of injections performed near a sensor can be many times higher.}
    \label{fig:composite}
\end{figure}

The composite coupling function computed via~\eref{eq:ccf} is used for comparing coupling between different sensor locations and producing estimates of DARM amplitudes, e.g. as part of event validation (see Section~\ref{sec:vetting}). Therefore we refer to a sensor's composite coupling function simply as its coupling function from here on. \Fref{fig:composite} provides an example of an estimated ambient for an accelerometer on the HAM6 vacuum chamber (which houses the interferometer output optics). The PEM website provides coupling functions for all accelerometers, microphones, and magnetometers produced from the most recent campaign of injections~\cite{PEM_website}.

\subsection{Uncertainties and Limitations}\label{sec:uncertainties}

We characterize coupling using the coupling function defined in \eref{eq:cf} instead of a transfer function because we do not assume perfect coherence. Low coherence can arise either due to non-linearity in the coupling or due to the spacing between the sensor and coupling site.

To measure coupling, we inject signals large enough to produce a response in DARM, but the maximum amplitude of injections is limited by the sensitive range of the environmental sensors (saturation produces an overestimate of coupling). This effectively limits how far below the DARM background we can probe for coupling or establish upper limits.

\Eref{eq:cf} relies on two assumptions about the coupling mechanism. First, the coupling is assumed to be linear, e.g. doubling the amplitude of the injection would double the amplitude of the response in DARM. We check this by repeating injections with different amplitudes. Second, the coupling function ignores any up- or down-conversion of the signal between the sensor and DARM. This non-linear coupling can be very significant for scattering noise and bilinear coupling but is not accounted for in the estimates of linear coupling. One way we detect non-linear coupling is by sweeping single frequency injections over time and searching for off-frequency response in DARM spectrograms.  Frequency changes from non-linear coupling can be an issue in broadband injections where up- or down-converted noise in DARM appears in the injection band, resulting in artificially higher estimates of linear coupling. We split broadband injections into smaller frequency bands to avoid this effect when necessary. One approach for quantifying non-linear coupling is presented in Washimi et al.\ (2020)~\cite{Washimi_2020}.

\subsubsection{Uncertainties associated with small numbers of sensors.}

As mentioned above, the use of \eref{eq:cf_full} relies on the assumption that the environment is monitored at the coupling site. The density of sensors is not great enough for this to be strictly true, especially if the source of the environmental signal is closer to the coupling site than the sensor is. The finite spacing of sensors leads to imperfect coupling functions but, for environmental signals that are generated at a distance greater than the typical sensor spacing of a few meters (the external signals that are the focus of PEM), the uncertainty can be estimated based on the differences in coupling measured at injections made at different locations. Given a sensor near a known coupling site, varying the injection location scales the signal amplitude as measured by the sensor and by DARM, in turn scaling the single-injection coupling function by some factor. Since the scaling is multiplicative, we quantify this variance by computing the geometric standard deviation of coupling factors in each bin.

\begin{figure}
    \centering
    \includegraphics[width=\textwidth]{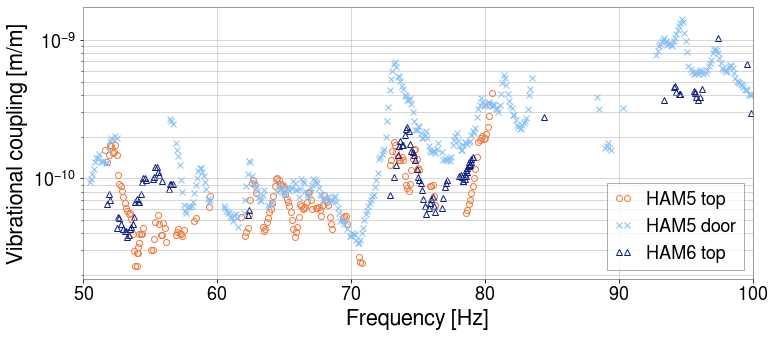}
    \caption{Single-injection coupling functions (upper limits not shown) for the HAM5 Y-axis accelerometer from shaking injections made from three different locations (on top of HAM5, on top of HAM6, and on the HAM5 chamber door) show the typical spread in coupling that results from varying the injection location. Multiple injections at different frequency bands are shown for each source location. On average the coupling measured from different locations varies by a 1.4.}
    \label{fig:injection_locations}
\end{figure}

\Fref{fig:injection_locations} shows single-injection coupling functions for an accelerometer measured from shaker injections produced from three locations (the distribution of injection locations is discussed in Section~\ref{sec:injections}. Since the injection locations are close enough to the accelerometer, we can assume that the variance is entirely due to the distance between the sensor and the coupling site. Averaged across all bins, the geometric standard deviation between injection locations is 1.4, i.e. coupling functions measured from vibrational injections, as well as vibrational noise projections to DARM such as those shown in \Fref{fig:ambient-vib}, vary by a factor of 1.4. A similar study combining geometric standard deviations for various magnetometers at both observatories shows that magnetic coupling measurements and noise projections vary by a factor of 1.7~\cite{cf_uncertainty}.

In the case of acoustic injections, the uncertainty in a coupling function can be exacerbated when nodes and anti-nodes in the acoustic signal coincide with the location of a sensor but not a coupling site. This results in peaks and troughs in the sensor spectrum at frequencies that have a node or anti-node at the sensor location, respectively. These artifacts can impact any sensor, but are more noticeable in microphone spectra than accelerometer spectra, possibly because the stiffness of the vacuum enclosure results in effectively averaging over a larger area; in microphones, the peak-to-trough ratio is typically a factor of a few. The peaks and troughs are present in the sensor but not in DARM, because the sensor monitors a single point whereas the coupling to DARM is spread across a large enough area for the effects of nodes and anti-nodes to average out. Consequently, this effect imprints troughs and peaks onto the coupling function.

The artifacts can be smoothed out of the spectra by computing a moving average over $X_{\mathrm{inj}}(f)$. The peak-to-peak distances are typically a few Hz, so we smooth the spectra enough to remove features up to a few Hz across. Since this can also result in less accurate coupling factors for sharp mechanical resonances, we use a moving average window that balances the smoothing of artifacts against the lost accuracy. The microphone spectra are smoothed with a logarithmically-scaled window which is set to 15\,Hz wide at 100\,Hz and 150\,Hz wide at 1000\,Hz.

\subsubsection{Testing coupling functions with off-site signals.}

Although the injections used to measure coupling functions are designed to best replicate environmental noise, there are still differences and it is useful to test the coupling functions with different environmental events by comparing noise seen in DARM during such events to noise levels predicted by PEM sensors and their coupling functions. Thunderstorms are known to produce short-duration transients in DARM at tens of Hz. At LLO, coupling functions for several accelerometers at the Y end station, where vibrational coupling was the highest, were capable of estimating the amplitude of multiple transients in DARM to within a factor two during a particularly loud thunderstorm~\cite{alog_thunder}. Helicopter flyovers can produce narrow-band features in DARM up to tens of seconds long. Coupling functions of various sensors at both interferometers predicted the amplitudes of lines produced by multiple helicopter flyovers during O3 to within a factor of two in most cases~\cite{alog_helicopter}. Vibrational noise from rain and the building heating, ventilation, and air conditioning (HVAC), which produce much longer-duration noise in DARM, have also been well estimated by coupling functions at LHO~\cite{alog_rain, alog_hvac_coupling}.

\section{Injection Methods}\label{sec:injections}

The basic methodology of environmental noise injections is described in~\cite{Effler_2015}. Here we summarize the methods and describe improvements made to the hardware and techniques since then.

\begin{figure}
    \centering
    \includegraphics[width=\textwidth]{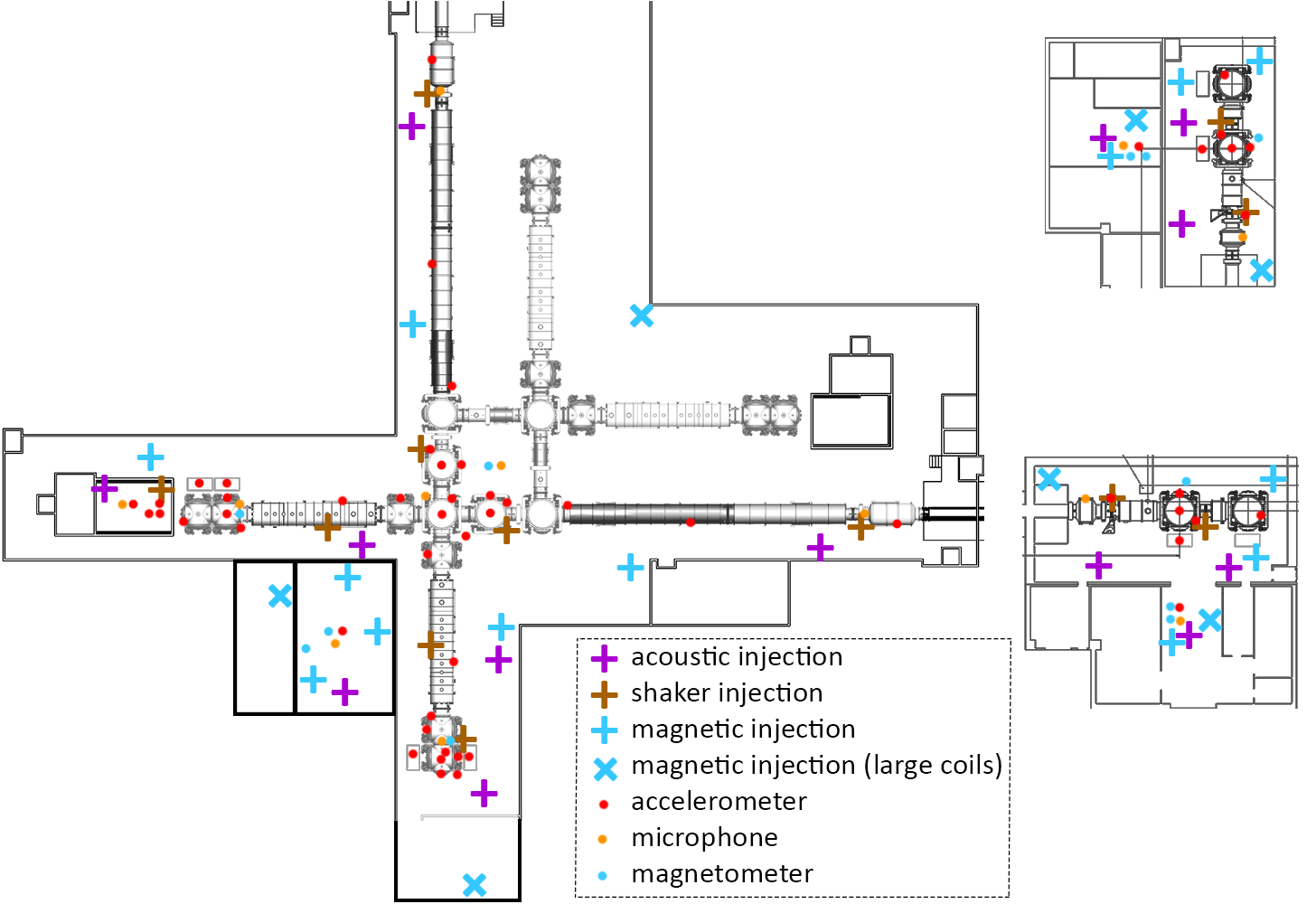}
    \caption{Standard locations for vibration and magnetic injections at the LHO corner station (left), Y end station (top right), and X end stations (bottom right). At each observatory we inject from 13 locations with acoustic injections, about 12 with shaking injections, and 15 with small-coil magnetic injections, with 7 large-coil magnetic injection locations planned for O4 (as explained in Section~\ref{sec:magnetic-injections}). The number and locations of shaker injections vary between injection campaigns. For all injection types, multiple injections are made at each location in order to focus on different frequency bands. Additionally, impulse injections (not shown) are made at locations where vibrational injections have revealed strong coupling sites.}
    \label{fig:injection-map}
\end{figure}

Injection locations are chosen to best mimic disturbances from outside the detector (\Fref{fig:injection-map}). To do so we choose them to be as far from the detector and environmental sensors as possible, but we are usually limited by the size of the detector sites themselves (some injections can be made from outside). 
Time dedicated to these tests has to be balanced against other instrumental work and observing time, which leads to a trade-off between measurement uncertainty and coverage. We perform injections from as many locations as time allows in order to maximize coverage of potential coupling sites. Increased time allocation towards environmental studies in recent years has allowed for a significant increase in the number of injection locations.

\begin{figure}
    \centering
    \includegraphics[width=\textwidth]{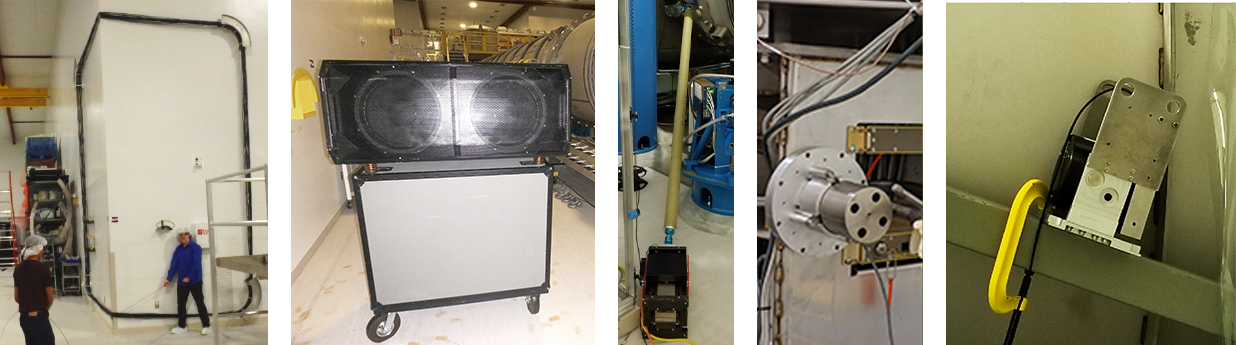}
    \caption{Injection equipment photos. From left to right: wall-mounted magnetic field injection coil; 14-in. speakers; APS 113 shaker connected to the door of a vacuum chamber by a rigid fiberglass rod; modified Piezosystem shaker clamped to an electronics rack; modified B\&K shaker clamped to a beam tube support.}
    \label{fig:injection_equipment}
\end{figure}

Table~\ref{tab:injectors} summarizes the current equipment used and \Fref{fig:injection_equipment} shows photos of some of the equipment. Seismic injections at low frequency (up to tens of Hz) during initial LIGO were performed with small electromagnetic and piezoelectric shakers~\cite{bk, piezo} and a weighted cart. A large shaker~\cite{big_shaker} has been used since the beginning of noise studies for O3. The large shaker can impart up to 133 N of sine force and a peak-to-peak displacement of 158 mm, compared to the electromagnetic shaker which imparts up to 45 N of force and a displacement of 8 mm.

\begin{table}
    \caption{\label{tab:injectors}Specifications for injection equipment.}
    \begin{indented}
    \item[]\begin{tabular}{@{}ll}
        \br
        Equipment & Injection type \\
        \mr
        Custom enclosure with two 14-in. speakers & Acoustic\\
        Various smaller speakers & Acoustic\\
        APS 113 Electro-Seis\textsuperscript{\textregistered} Long Stroke Shaker~\cite{big_shaker} & Vibrational\\
        Piezosystem\textsuperscript{\textregistered}~\cite{piezo} shaker with custom reaction mass & Vibrational\\
	    Br\"uel \& Kj\ae r\textsuperscript{\textregistered}~\cite{bk} EM shaker with custom reaction mass & Vibrational\\
	    1\,m diameter copper coil (100\,turns) & Magnetic\\
	    3\,x\,3\,m and 5\,x\,5\,m coils (80-100\,turns) & Magnetic\\
        \br
    \end{tabular}
    \end{indented}
\end{table}

\subsection{Vibrational injections}\label{sec:vibrational-injections}

Two new injection techniques have been developed for localizing vibration coupling sites connected to the vacuum enclosure, such as locations on the vacuum enclosure that reflect scattered light. The techniques rely on the slow propagation speeds (hundreds of meters per second) of vibrations on the steel vacuum enclosure walls or, for acoustic injections, in air. These two techniques aided in the localization of a coupling site that was producing a 48\,Hz peak in DARM throughout the first half of O3, as discussed in Section~\ref{sec:scattering}.

\begin{figure}
    \centering
    \includegraphics[width=\textwidth]{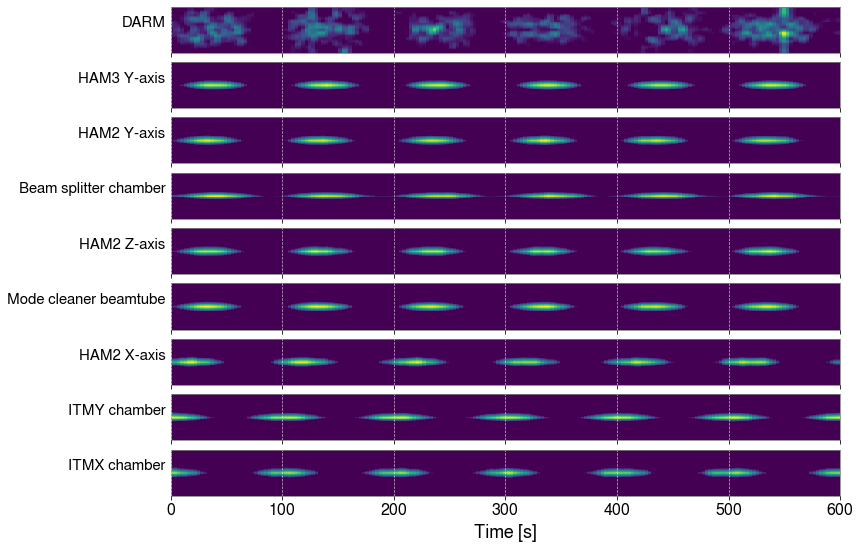}
    \caption{Example spectrograms showing a beating-shakers injection using two shakers to localize the coupling site responsible for a 48\,Hz noise peak in the DARM spectrum. The shakers were injecting at 48 and 48.01\,Hz. The Y-axes of the spectrograms are centered along at 48\,Hz and show the combined signal in each sensor modulating at the beat frequency (0.01\,Hz). This set of spectrograms suggests that the accelerometers on the input test mass (ITM) chambers and the Y-axis HAM2 accelerometer are likely not close to the true coupling location, since the beat envelopes are the furthest offset from the beat envelope in the DARM response. Multiple other injections were made (not shown here) with varying shaker locations in order to rule out other sensors until the most likely candidate remaining was the HAM3 Y-axis accelerometer. Black glass was used to block scattered light at this location and the peak was eliminated for the second half of the O3 observation run.}
    \label{fig:beats}
\end{figure}

The beating-shakers technique is narrow-band, and involves vibrating the vacuum enclosure at two slightly different frequencies, each injected from a shaker or a speaker at a different location (e.g. a shaker at one location injects a sine wave at frequency $f$ and a shaker at the other location injections at frequency $f + 0.01$\,Hz). The two injections are adjusted in amplitude to produce strong beats in DARM. Because the injection locations are different, the relative phase of the two injected signals varies with location on the vacuum enclosure. As a result, the phase of the beat envelope varies with position, and different sites experience maximum chamber wall motion at different times. The sites with accelerometer signals that have the same beat envelope phase as DARM are candidates for the scattering sites on the vacuum enclosure walls (\Fref{fig:beats}). Other sensors that are not near the coupling site may also match the phase by chance, but these false positives can be rejected by varying the locations of the shakers.

\begin{figure}
    \centering
    \includegraphics[width=0.9\textwidth]{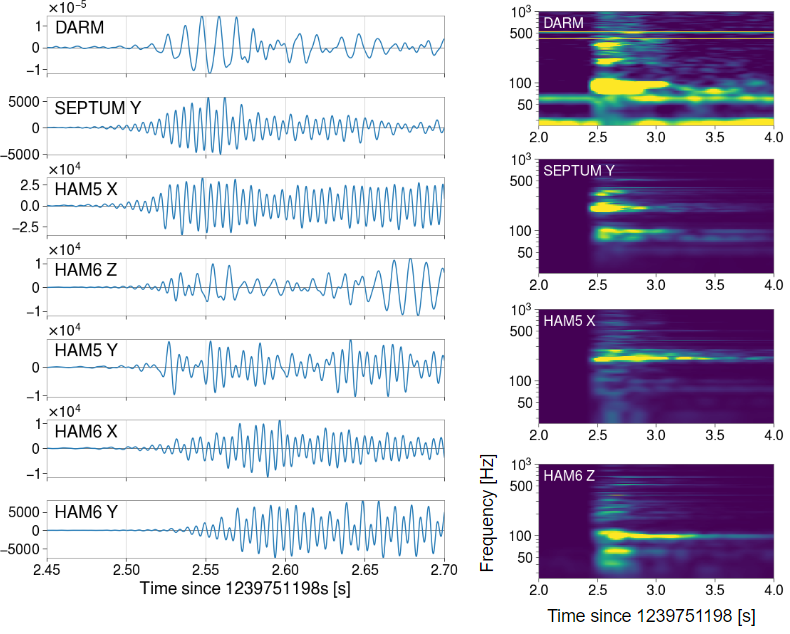}
    \caption{Left: Example time series of a single impulse injection signal in DARM and various output optics accelerometers. Multiple sensors observe an impulse time-of-arrival matching that of DARM, but repeating the injection from various other locations rules out sensors that do not match DARM consistently across multiple injections. In this case the septum (separating the HAM5 and HAM6 chambers) accelerometer signal matched the DARM signal most consistently (other injections not shown for brevity). Right: Spectrograms of the same impulse injection for DARM and the three sensors with the closest matching time-of-arrival to DARM. The similarity between the frequency structure of the septum accelerometer and that of DARM further supports the septum as a dominant coupling site in the output arm.}
    \label{fig:impulse}
\end{figure}

The second injection technique, which is broad band, involves propagation delays in impulse injections. Impulse injections are performed by striking the vacuum enclosure directly with enough force to produce a transient in DARM and in nearby accelerometers. The vibrational impulse propagates through the structure of the vacuum enclosure, arriving at different accelerometers and coupling sites at different times. We can distinguish these arrival times because the propagation velocity is much slower than in solid material, and is only roughly 300\,m/s in our case. Using time series plots, the arrival time of the impulse in DARM is compared to the arrival time of the impulse in multiple accelerometers (\Fref{fig:impulse}, left). The accelerometers that have the same arrival time as DARM are more likely to be near a coupling site than those that observe the impulse much earlier or later than DARM does. Again, varying the location of the injection eliminates sensors that match the DARM time-of-arrival by chance but are actually far from the coupling site. An additional consistency check is that the coupling of accelerometers near the coupling site will vary less between different impulse locations than that of accelerometers far from the coupling site. Finally, if the accelerometer is at the coupling site, the impulse in DARM will have a resonance structure that is similar to the resonance structure of the accelerometer signal, which can be judged from spectrograms (\Fref{fig:impulse}, right).

\subsection{Magnetic injections}\label{sec:magnetic-injections}

Improvements have also been made to the magnetic field injection equipment. In order to generate fields strong enough to couple into DARM using the 1\,m magnetic field coils made during initial LIGO~\cite{Effler_2015}, we must focus the power of the coil into narrow bands and combs instead of injecting broadband signals. This was sufficient in initial LIGO when strong magnetic coupling occurred primarily through permanent magnets. However, due to the removal of permanent magnets from the test masses, coupling from those sources has decreased and cables and connectors have become the dominant coupling sites above about 80\,Hz, introducing more structure to the coupling functions and requiring stronger injections.

\begin{figure}
    \centering
    \includegraphics[width=0.8\textwidth]{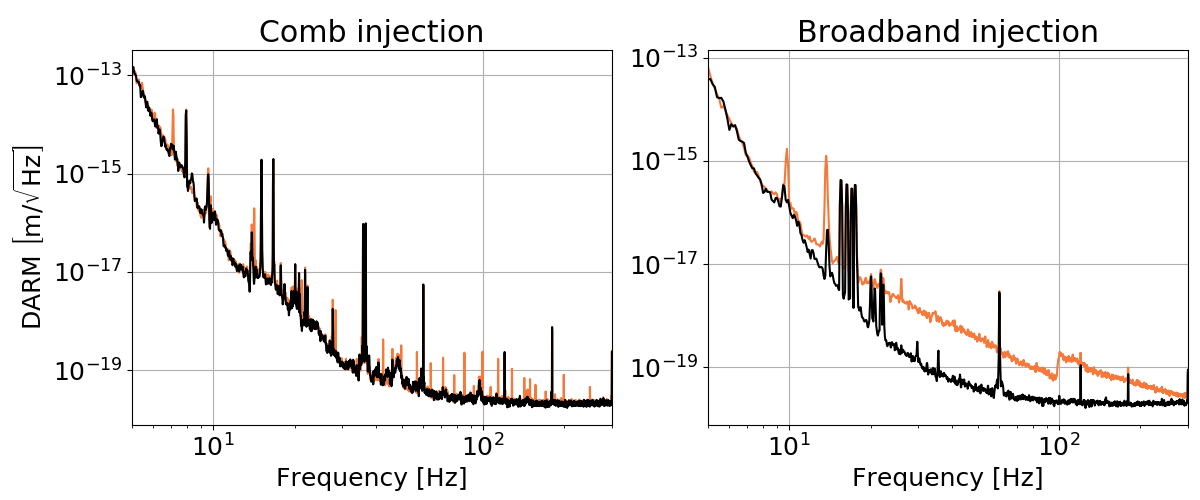}
    \caption{DARM response to old (left, comb) and new (right, broadband) magnetic injections. Black and orange lines show the DARM ASD before and during the injections, respectively. The comb injection curve is a composite of multiple comb injections, with fundamental frequencies of 7.1\,Hz, 14.2\,Hz, and 49.7\,Hz, made at different times. The broadband injection spectrum is a composite of a 10-100\,Hz injection and a 100-1000\,Hz injection made at different times, hence the break at 100\,Hz.}
    \label{fig:bigcoil}
\end{figure}

To achieve high-amplitude broadband magnetic injections, seven wall-mounted coils, each one a 3\,m x 3\,m or 5\,m x 5\,m square of 80-100\,turns, are being installed at each site; three at the corner station and two at each end station. These coils are fixed in place and can be operated remotely, allowing for weekly injections to monitor variations in magnetic coupling caused by changes to electronics. \Fref{fig:bigcoil} compares the old and new magnetic injections. Some coils were installed and operated at the sites during O3; the project will be completed by the start of O4.

\section{Mitigation of Environmental Effects in aLIGO}\label{sec:O3}

We use the methods discussed thus far to track down  noise sources whose estimated ambient level in DARM is more than a tenth of the DARM background. Mitigation can be accomplished in three ways: by removing or modifying the source itself, by isolating the source or otherwise addressing propagation of the signal to the detector, or by reducing the coupling itself through some modification to the detector. Here we provide several examples of environmental effects that were mitigated based on results from noise investigations.

\subsection{Seismic and acoustic influences}

\begin{figure}
    \centering
    \includegraphics[width=0.8\textwidth]{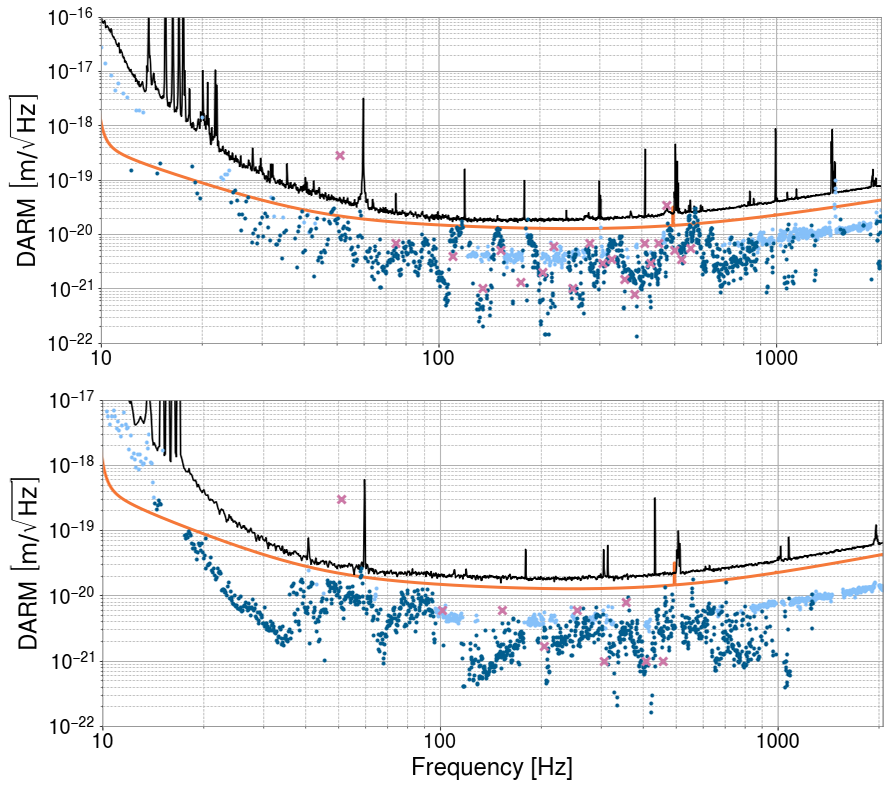}
    \caption{Ambient vibrational noise at LHO (top) and LLO (bottom), shown in dark blue (measurements) and light blue (upper limits). The values are produced by selecting the highest-amplitude composite coupling function at each bin across all sensors at each observatory. The black and orange lines show the DARM background and the aLIGO design sensitivity, respectively. Purple crosses show the initial LIGO ambient from S6. The comparison between the S6 and O3 ambient is complicated by the dramatic hardware differences between the initial LIGO and aLIGO detectors. The dependence of coupling on specific hardware details is exemplified by the differences between the O3 ambient noise levels in LHO and LLO, which are designed to be identical.}
    \label{fig:ambient-vib}
\end{figure}

\Fref{fig:ambient-vib} shows the ambient contribution of vibrational noise during O3, produced by combining the highest coupling factors among accelerometers and microphones measured from an injection campaign at the beginning of O3. At the end of O3,  the vibration noise background at both observatories was dominated by input beam jitter above 100\,Hz (discussed in Section~\ref{sec:jitter}). At LHO,  the dominant coupling region below 100\,Hz was the output arm. At LLO, the dominant coupling regions were the Y-end in the 40-60\,Hz band and the output arm in the 60-100\,Hz band.

\subsubsection{Input beam jitter.}\label{sec:jitter}

Alignment fluctuations of the beam (beam jitter) entering the interferometer cause variation in the coupling of the fundamental mode into the arm cavities, producing amplitude noise. In addition, the varying beam position relative to defects in the test masses causes a variation in the balance of light between the two arms, further contributing to the noise~\cite{Buikema_2020}. Beam jitter turned out to be the source of large peaks that appeared in DARM at the beginning of aLIGO, that were well predicted by PEM coupling functions for laser enclosure sensors~\cite{Schofield_2015}. The dominant source of alignment fluctuations was turbulence in the laser cooling system, causing vibration of mirrors and other optics on the table and in the laser, producing peaks in DARM at mechanical resonances of the optics and their mounts. A second mechanism may be variation in beam diameter associated with the turbulent cooling. In addition to vibrations from the cooling system, transient vibrations, such as those made by large vehicles or heavy footsteps in the control room, produced transients in DARM by temporarily increasing alignment fluctuations.

Mitigation has included removing turbulence-producing connectors, sharp turns in the coolant lines, and abrupt diameter changes within the cooling system; and reducing the flow of coolant (replacing "quick-connects" reduced table motion by a factor of about 2)~\cite{alog_psl_chiller}. In addition, injections were used to identify the optic mounts that produced the largest peaks in DARM, and mass-and-viton dampers were added to the mounts. This resulted in motion reductions by factors of a few~\cite{alog_optic_damping}. Finally, the resonances of optic mounts on the periscope that raises the beam from the laser table level to the interferometer level, were tuned by adding small masses to the optic mounts, so that their resonances would not overlap in frequency with periscope resonances that would increase the motion of the mounts~\cite{alog_periscope}.

\subsubsection{Vibrational coupling at resonances of the vibration isolation system.}

One of the most troubling environmental couplings early in aLIGO was vibration coupling at 100\,Hz and above through the seismic isolation system. Not only were ambient vibration levels producing noise within a factor of two of the DARM background around 1000\,Hz, but the coupling was highly non-linear (see non-linearity discussion in Section~\ref{sec:uncertainties}), and it was the only vibrational coupling observed that could produce noise in the detection frequency band around 100\,Hz  from a source near 1000\,Hz.  One could imagine a rising frequency signal (chirp) from the startup of a motor with squealing bearings, for example, that would have been able to produce a chirp in DARM in the 100\,Hz band. This problem required special vetting of the first GW detections since, normally, the vetting procedure only assumes linear coupling (discussed further in Section~\ref{sec:vetting}).
 
The coupling was due to little or no isolation in certain frequency bands associated with mechanical resonances of the isolation system. The active system vibrationally isolating the in-vacuum optical tables works mainly below 20\,Hz. For higher frequencies, there are one (HAM chambers) or two (BSC chambers) passive isolation layers associated with the suspension of the optical tables. But, at the many resonances (violin modes) of the multiple wire-like flexures that suspend the tables, there was little isolation, allowing vibration at these frequencies to couple to DARM. This lack of isolation produced linear vibration coupling at multiple optical tables and, at the dark port table, non-linear coupling due to an intermodulation of vibration and a strong length dither used in controlling the length of the output mode cleaner (OMC). The coupling was mitigated by attaching Viton\textsuperscript{TM}~\cite{viton} to the suspension flexures at tables with coupling to DARM. This reduced the linear and non-linear coupling by a factor of about three~\cite{alog_HAM6}. To further reduce non-linear coupling, the amplitude of the OMC length dither was reduced as far as possible, yielding a further reduction in non-linear coupling by a factor of eight~\cite{alog_HAM6}.

\subsubsection{Coupling of wind through ground tilting in the 0.1 Hz band.}

Vibrations from wind affect the interferometer directly in the 10-100\,Hz band. At lower frequencies, particularly in the band around 0.1\,Hz, pressure fluctuations associated with wind can affect sensitivity and duty cycle by tilting the ground. Interferometer performance can be affected by direct tilt of optical table supports or by tilt of ground motion sensors used in the active isolation system, producing inaccurate signals from sensors that do not distinguish between tilt and acceleration. Even far from the buildings, we found that the ground tilts in wind (about $1 \times 10^{-8}\ \mathrm{rad}/\sqrt{\mathrm{Hz}}$ at 0.1\,Hz in wind reaching 15\,m/s at LHO), to a degree that is consistent with spatially varying wind speeds and Bernoulli forces. But the tilting in the buildings was a factor of several times larger, and found to be greatest near the building walls. The pressure fluctuations on the walls are thought to tilt the wall supports which, in turn, tilt the ground at their base. The coherence length of floor tilt measured at Hanford was a couple of meters, indicating that the cement slab does not tilt as a unit. Instead, the tilting is local and mainly within meters of the base of columns that support the walls, consistent with an elastic dimpling of the ground around the support~\cite{alog_tilt_1, alog_tilt_2}.

The localized nature of the dominant tilt has led to the simple mitigation technique of moving ground sensors as far from the walls as possible. While certain sensors could be moved, the large vacuum chambers near the wall could not, and for future installations, we have recommended that the chambers be placed at least 10\,m from the base of wall supports.

In order to further mitigate the effects of wind-induced tilt, tilt meters with improved sensitivity were designed and deployed~\cite{Venkateswara_2014, Venkateswara_2017}. The first versions were produced to correct the artifacts that tilts produce in seismometers, but a table-top tilt meter is also being developed in order to mitigate the effects of the tilt of the optical tables in the chambers. 

Wind fences have been used to reduce wind in agricultural and recreational settings, and modeling suggested that wind fences may be useful for reducing the effects of wind pressure on the building walls. For this reason a wind fence was constructed at Hanford, and is currently being evaluated~\cite{wind_fence}. One remaining question is how effective a wind fence is in the troubling frequency band around 0.1\,Hz where the length scale is 100\,m for 10\,m/s wind.

\subsubsection{Vibration modulation of scattered light paths.}\label{sec:scattering}

A major source of detector noise and reduced sensitivity to GWs is the scattering of light from the beam spot on a test mass or other optic to surfaces that are moving relative to the optic, like vacuum chamber walls. A very small fraction of the light reaching the moving surface is reflected to the originating or another beam spot, where it scatters back into the main interferometer beam. As the distance to the moving surface changes, the phase of the returning light changes relative to the main beam, producing fluctuations in the amplitude of the beam, that, at 1 part in $10^{20}$ can be on the scale of those produced by gravitational waves. In addition to this  sensitivity to recombined scattered light, the scattering noise is problematic because of non-linear coupling when the path length modulation becomes comparable to the wavelength of the light, producing noise at harmonics of modulation frequencies~\cite{Soni_2020}.

\begin{figure}
    \centering
    \includegraphics[width=0.8\textwidth]{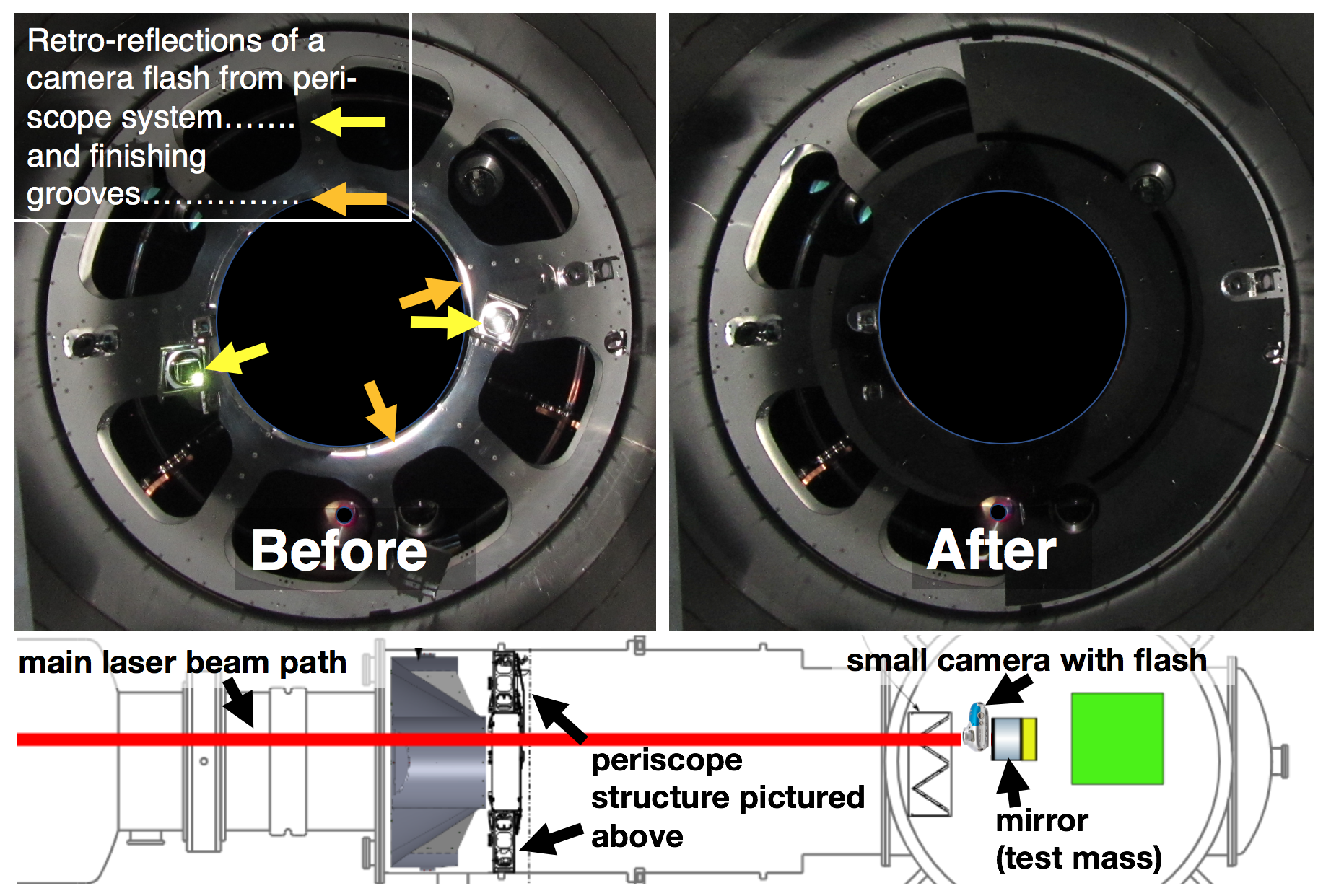}
    \caption{Diagnostic photographs taken from the point of view of a test mass beam spot, before and after scattering noise mitigation. The image on the left shows retro-reflections of light from the camera flash, which follows a similar path as light scattered from the interferometer beam by imperfections in the test mass. The photograph was used to identify potential sources of scattering noise. The structure is a calibration periscope in the beam tube. The photograph on the right was taken after removal of the mirrors and after the subsequent installation of baffling, both of which reduced the scattering noise in the GW channel. The central black disk was added to the image to avoid confusion from light reflected by a gate valve that is withdrawn during operation.}
    \label{fig:scattering}
\end{figure}

The subtlety of scattered light noise is illustrated by the mechanism that was behind a mysterious glitch in DARM that turned out to be produced by ravens~\cite{alog_ravens, Nuttall_2018}. The Rube Goldberg–like mechanism began in the desert sun at LHO, where ravens pecked at ice accumulations on a cryopump vent tube just outside of an end station building. The vibrations from pecking were transmitted through the vent tubes to the cryopump inside the building. The cryopump was attached to the beam tube, and the vibrations were transmitted through the beam tube to a calibration structure located inside of the vacuum, which vibrated slightly with each peck. The structure was angled so as not to retro-reflect light scattered from the test mass, about 10\,m away. However, polishing grooves on the surface reflected a small fraction of the light back to the test mass, where a small fraction recombined with the main beam. The interference between the light in the main beam and the tiny amount of light reflected from the grooves varied with the motion of the calibration structure produced by each peck. The varying interference caused fluctuations in the light of the main beam, similar to the fluctuations produced by gravitational waves. After the discovery of this coupling mechanism, the calibration structure was baffled to reduce the light scattered back into the interferometer, eliminating the raven glitches and other similar vibrational signals~(\Fref{fig:scattering}).

Diagnostic photographs can be used to identify a common type of scattering path that involves light that is scattered from the beam spot on an optic to a moving reflective surface and back to the beam spot, where it scatters back into the main beam (\Fref{fig:scattering}). Problematic reflective surfaces often depend strongly on precise angles and surface finish, and they can be difficult to identify in design drawings. To find potential reflective surfaces, during incursions such as optic installation, we place a small camera (with the camera flash very near its aperture) as close to the face of an optic as possible, and look for bright reflections of the flash in photographs taken from the optic's point of view. Most metal surfaces that would directly reflect infrared light scattered from the face of the optic also directly reflect camera flashes from the face of the optic. Since reflections are common and it is difficult to fix each one, we have begun work on a system to roughly rank the noise potential of reflective surfaces in the photographs, using the estimated coupling of scattered light to the GW signal at the particular optic, the  solid angle of the reflecting surface and its distance from the optic, the approximate motion of the surface, and the estimated angular dependence of the scattering from the optic along with the angle to the reflector~\cite{alog_reflections}.

\begin{figure}
    \centering
    \includegraphics[width=\textwidth]{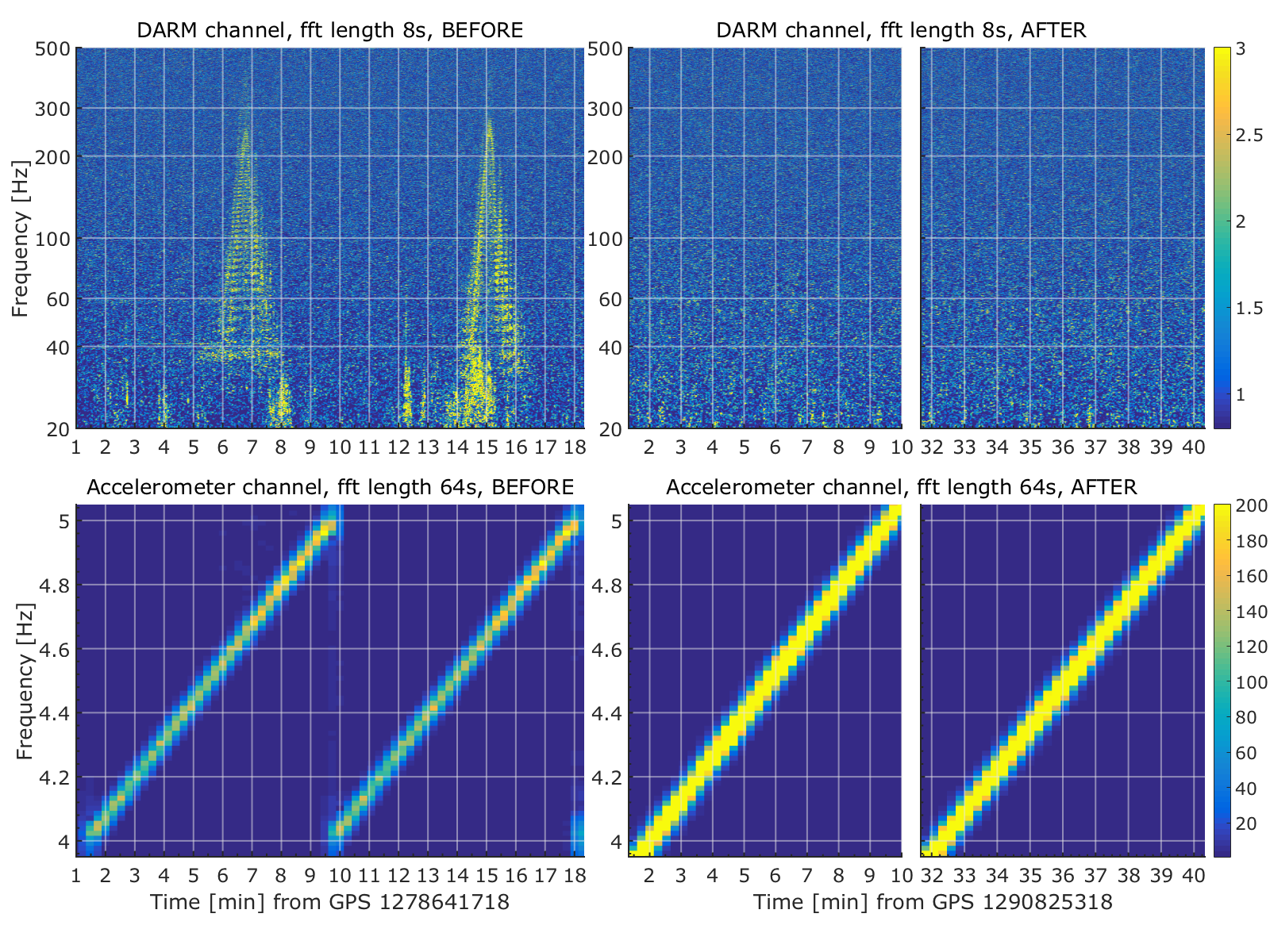}
    \caption{Mitigation of high-motion cryo-baffle scattering noise. Spectrograms of the DARM channel (top) and an accelerometer channel (bottom) showing consecutive vibrational frequency sweep injections that excite noise in DARM as they cross 4.62\,Hz before mitigation. The ``AFTER'' sweep is composed of two graphs such that it shows only the relevant range of the sweep to be compared with ``BEFORE''. The spectrograms are normalized by median, but the spectra are not very different between the two tests such that they are directly comparable. The Q was determined to be around 450 and so it is easy to get to high velocities by exciting this resonance - identified on a baffle at the end station. This resonance was damped in between the two measurement sets, such that the same sweep with even higher amplitude no longer makes any noise in DARM.}
    \label{fig:sweep}
\end{figure}

Scattered light baffles can themselves be problematic - for example, vibrations in the 5-30\,Hz band, such as from nearby truck traffic, produced transient noise and limited detector sensitivity in early aLIGO. Investigations using vibration injections and laser vibrometry showed that the coupling was due to light reflecting from imperfect light baffles. The ground motion was amplified by the resonances of the baffles (quality factors of several hundred), increasing the velocity by several hundred times, and producing scattering noise that reached hundreds of Hz for 10\,Hz excitations~\cite{alog_swisscheese1}. The problem was solved by damping the baffle resonances with Viton\textsuperscript{TM} in order to reduce their velocity~\cite{alog_swisscheese2}. Eventually, the most reflective parts of the baffles were also removed. Near the end of O3,  a second type of un-damped baffle was identified as a noise source~\cite{alog_cryobafflemovie, alog_cryobaffle_llo1, alog_cryobaffle_lho} and damped~\cite{alog_cryobaffle_llo2}~(\Fref{fig:sweep}).

\begin{figure}
    \centering
    \includegraphics[width=\textwidth]{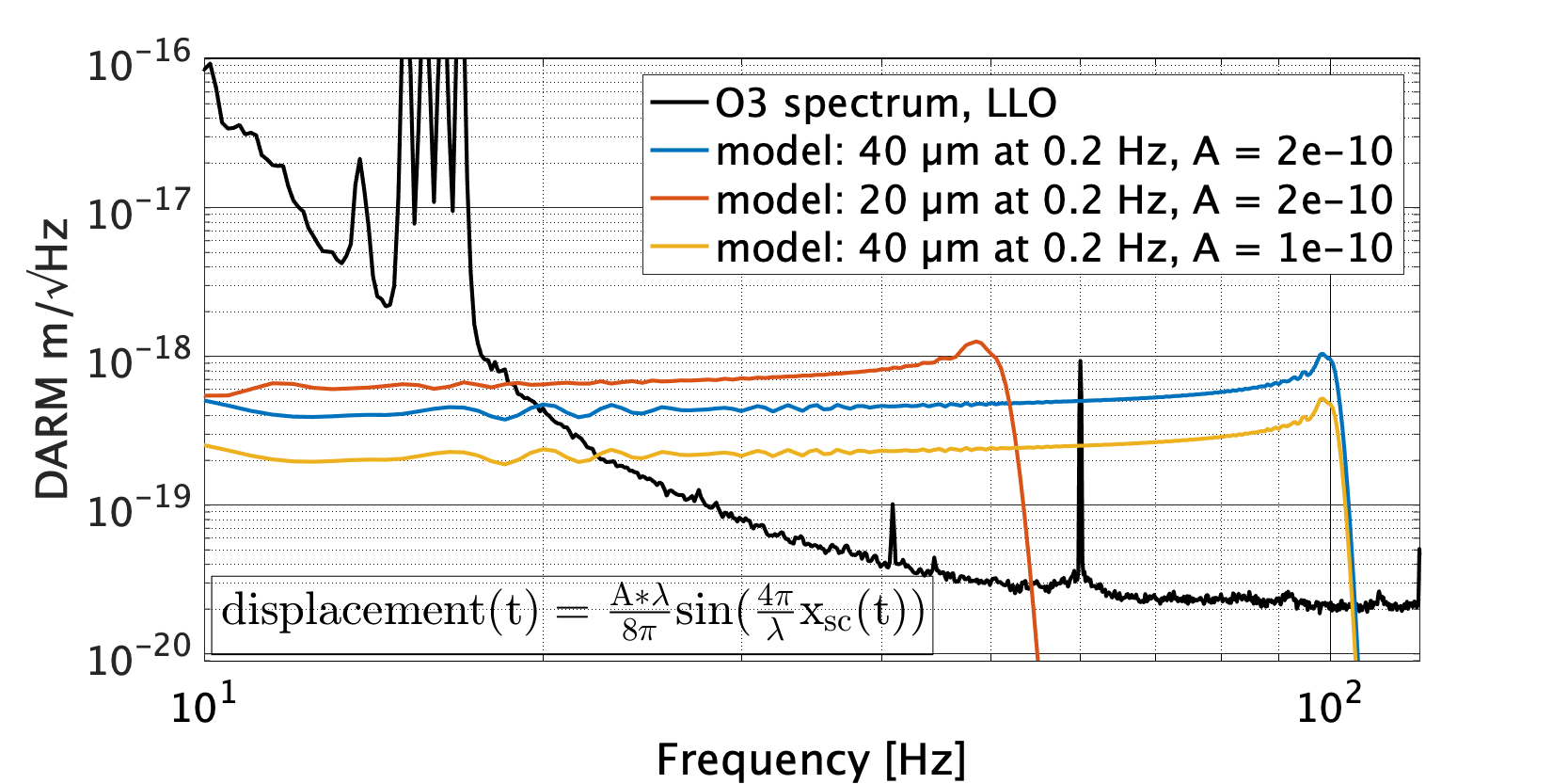}
    \caption{Modeled mitigation of high-motion scattering noise (blue) by reducing the velocity of the reflecting surface (orange) or the amplitude of the reflected light (yellow). The model is based on the inset formula~\cite{Ottaway_2012} where: A is the amplitude of the scattered light field returning to the main beam, $\lambda$ is the wavelength of the light, and $x_{\textrm{sc}}(t)$ is the motion of the scattering surface. Note that as the motion is reduced, for the same light field amplitude the resulting displacement increases as the square root of the motion reduction factor. The modeled motion is a sine wave with frequency of 0.2\,Hz, which simulates the largest ground motion at the two sites due to ocean waves. }
    \label{fig:shelf}
\end{figure}

Damping is an effective tool for reducing scattering noise because it lowers the velocity of the reflector. Scattering noise can be mitigated by reducing the amplitude of the scattered light or by reducing the velocity of the reflector, or both. Reducing the amplitude by a factor of two yields a factor of two reduction in noise. Reducing the velocity by a factor of two, such as by damping, reduces the maximum frequency of high-motion scattering "shelves" by two, and, above that frequency, the improvement can be greater than a factor of two (\Fref{fig:shelf}).  

An important driver of scattering noise for long interferometers is ground motion at the ocean-wave driven microseismic peak frequency, in the 0.1-0.3\,Hz region, which has produced noise in DARM that reached nearly 100\,Hz, several hundred times higher in frequency. Because the ground moves differently at the ends of the 4-km-long cavities, the control systems that minimizes relative motion of the test masses at opposite ends of the cavity can lead to micron-scale relative motion between the end test mass (ETM) and other objects in its vicinity that are not moved with the test mass and cavity. A major source of transients during the first three observing runs, especially when microseismic noise was high, was light reflected from the gold electrostatic actuator traces on the reaction mass behind the test mass. The variation in the optical path length was amplified by multiple reflections between the traces and the back of the reflective surface of the test mass~\cite{alog_microseism}. The problem was solved by driving the reaction mass to minimize relative motion between it and the test mass~\cite{Soni_2020}.

\begin{figure}
    \centering
    \includegraphics[width=\textwidth]{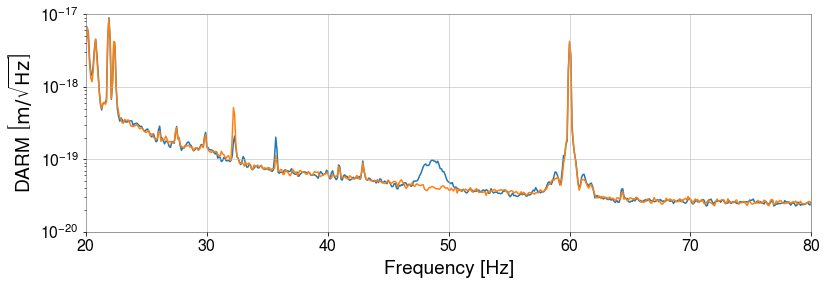}
    \caption{Mitigation of the 48\,Hz peak caused by scattering noise from the HAM3 vacuum chamber; DARM spectrum before (blue) and after (orange). }
    \label{fig:48Hz}
\end{figure}

In the reaction mass and baffle examples discussed above, and illustrated in Figures~\ref{fig:shelf} and~\ref{fig:sweep}, the variation in the optical path length of the scattered light was large compared to the wavelength of the scattered light. This results in "shelf" features like those in \Fref{fig:shelf}. A good example of low-motion scattering noise, where the optical path length variation is small compared  to the wavelength of the laser, is the 48\,Hz noise peak that was investigated using the techniques of Section~\ref{sec:vibrational-injections}. Impulse injections pointed to the highest coupling being near the vertex and input arm in the corner station. Using the beating-shakers injection method, with frequencies of 48 and 48.01\,Hz, multiple shaker pair locations and extra accelerometer instrumentation, it was found that the timing of the beat envelope in DARM best matched that of accelerometers that were mounted on the HAM3 door. The moving reflector was discovered to be in a vacuum chamber illuminator mounted on a view port in the HAM3 door, that had a surface normal to the incident scattered light~\cite{alog_48Hz}. \Fref{fig:48Hz} shows the improvement in DARM made by blocking the path to the illuminator with black glass.

\subsection{Magnetic influences}

\begin{figure}
    \centering
    \includegraphics[width=0.8\textwidth]{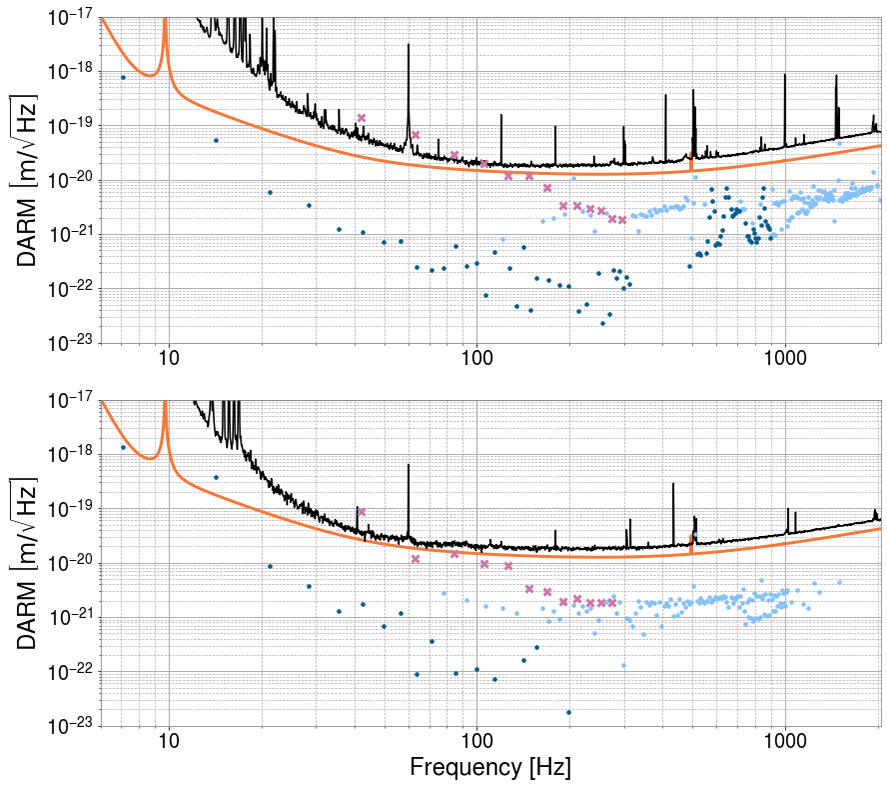}
    \caption{Ambient magnetic noise at LHO (top) and LLO (bottom), shown in dark blue (measurements) and light blue (upper limits). The values are produced by selecting the highest-amplitude composite coupling function at each bin across all sensors at each observatory. Only values at multiples of 7.1\,Hz are used because most of the injections performed were 7.1\,Hz comb injections (off-comb bins had too few injections to produce reliable composite coupling functions). The black and orange lines show the DARM background and the aLIGO design sensitivity, respectively. Purple crosses show the initial LIGO ambient from S6. The difference between the S6 and O3 ambient noise levels is mostly due to the removal of magnets from the test masses as described in Section~\ref{sec:detector_upgrades}.}
    \label{fig:ambient-mag}
\end{figure}

Magnetic injections early in aLIGO suggested that coupling to permanent magnets in the suspension system  could prevent LIGO from reaching design sensitivity in the 10-20\,Hz regions~\cite{Schofield_2013}. While the test mass actuator is electrostatic and not magnetic (as in initial LIGO), a number of permanent magnets were used in the suspensions, including for actuation in the first three of the four levels of the isolation chain and for eddy current damping. The greatest number of permanent magnets were in the eddy current damping arrays and these were removed. Nevertheless, ambient fields are still predicted to produce noise at greater than one-tenth of the design sensitivity in the 10-20\,Hz band (\Fref{fig:ambient-mag}), and may need to be further addressed as we reach design sensitivity in the 10\,Hz region.

At higher frequencies, generally above about 30\,Hz, the dominant magnetic coupling appears to be through induction of currents in cables and at connectors, mainly to actuator cabling and other cabling in the control system. Mitigation of coupling to cables and connectors has required a continuing program of monitoring coupling since cables are often disconnected and reconnected during runs as electronics are replaced for problems or upgrades. This program consists of making weekly, broadband magnetic field injections using the large wall-mounted coils described in~\ref{sec:injections}. The injections have shown that peaks can appear or disappear, as well as shift in frequency, on a weekly basis.

As explained in Section~\ref{sec:pem_sensors}, magnetometers can be used to identify sources of persistent spectral artifacts in DARM, even when the coupling mechanism is not necessarily through magnetic fields. Many examples of lines and combs mitigated throughout O1 and O2 through magnetometer studies are provided in Covas et al.\ (2018)~\cite{Covas_2018}.

While ambient fields do not normally limit the sensitivity of the interferometers for most astrophysical sources, the stochastic GW searches reach higher strain sensitivities by integrating data over multi-month periods and searching for correlations between sites. Magnetic fields that are correlated between sites could limit this sensitivity or even lead to misinterpretations of the GW background. Geomagnetic phenomena, such as the Schumann resonances discussed in Section~\ref{sec:pem_sensors}, could produce such correlations~\cite{Thrane:2013npa, Thrane:2014yza, Coughlin_2018}. In order to monitor such correlated fields, we have installed sensitive magnetometers far from the much greater uncorrelated magnetic fields in the buildings.

\subsection{Other influences}

\subsubsection{Dust and particulates.}

In initial LIGO, the photodiodes at the interferometer output were external to the vacuum system and glitches were produced in DARM by dust passing through the beams at the photodiodes, generally where the beams were smallest in diameter. For aLIGO, the photodiodes were placed in vacuum, but the main laser was not. The room containing the main laser is a clean room and we have not found dust to be a problem. However, excess vibration of the beam tube and vacuum chambers can cause oxidized metal particulate from the inside of the vacuum system to drop through the beams and produce glitches in DARM. An absence of a correlation between glitches in DARM and ground motion suggests that this problem is minimal, though it has shown up strongly when the beam tubes were being cleaned and when the vacuum chambers are struck~\cite{alog_beamtube}. For the beam tube, we estimated that particulate glitches were unlikely for accelerations less than 1\,m/s$^2$~\cite{alog_particulate_glitch}, which is still a concern because it could be reached in stick-slip events associated with thermal expansion of the beam tube. 

\subsubsection{Radio-frequency fields.}

Injections have shown that out-of-band RF fields (above 10\,kHz) have no influence on the interferometer until they reach amplitudes that are orders of magnitude above most of the background from external sources~\cite{alog_rf}. The strongest RF coupling was found to be at the 9 and 45\,MHz modulation frequencies. We monitor these frequencies and we scan for signals between 9\,kHz and 1\,GHz. 

Unlike externally generated RF fields, RF signals generated by the detector have been observed to produce noise in the gravitational wave channel, glitches called RF whistles~\cite{Nuttall_2015} and other features. For example, during the first observing run, a narrow-band feature intermittently appeared in the LHO DARM spectrum around 650\,Hz, resulting in spurious GW event candidates near that frequency~\cite{Palamos_2020}. The feature was found to be correlated with a similar feature observed by the RF receiver. The receiver actually witnessed the RF noise during a particular step of the interferometer's lock acquisition process, the transition to DC readout. This eventually led to discovering that the origin of the noise was the frequency of a voltage-controlled oscillator (VCO) used in lock acquisition beating with some other RF source in the inteferometer. Changing the frequency setting of the VCO eliminated the 650\,Hz line from DARM~\cite{alog_650Hz}.

\subsubsection{HVAC and other temperature control systems.}

A dominant source of vibration in the 1-10\,Hz band has been air and cooling water flow associated with the large central HVAC systems in the corner and end stations. Turbulent eddies in the air plenum downstream of the HVAC turbines were shown to produce vibrations that affected the interferometer. The large eddies producing pressure fluctuations in the few Hz region were broken up by installing screens in the outlets of the turbines, reducing ground vibration in the band around 5\,Hz.

Water chillers and pumps used to chill the HVAC system are generally isolated on springs, but turbulence in the pipes connecting the chillers to the air handling system was observed to increase ground motion in the 5-15\,Hz region. The turbulence was partially mitigated by reducing the chilled water flow using variable frequency drives to slow the water pumps. Isolation of the pipes from contact with the ground and larger diameter pipes could reduce this problem in future installations.

Temperature fluctuations, while generally in the mHz band, can affect interferometer performance by, for example, changing the length of the blade springs at the top of the pendulum suspensions, which lowers or raises the entire test mass suspension, resulting in rubbing. Building temperature control often employs sensors mounted on the walls. When these walls are external, the temperature in the region of the vacuum chambers may increase in cold weather and decrease in warm weather because the system attempts to maintain a constant temperature at the wall sensors. Sensors have been moved off of the walls and closer to the vacuum chambers to mitigate this problem~\cite{alog_temperature}.

In addition to centralized temperature control systems, vibrations from local cooling, such as electronics cooling fans, can produce noise in DARM. We have mitigated acoustic coupling by placing electronics racks in separate rooms from other interferometer components, and by seismically isolating the electronics racks on elastic legs. Cooling systems for lasers etc.\ have also been problematic and have required additional seismic isolation. 

\subsubsection{Other site activities.}

Vehicle movements on sites have produced transients in DARM. This often occurs as vehicles cross bumps and gravel. Mitigation has included deeper burial of pipes that cross roads, removal of gravel from roads, patching of cracks and prohibition of travel in certain areas during observation. Even heavy steps in a control room can couple by producing beam jitter (see Section~\ref{sec:jitter}).

The springs isolating motor-driven equipment have been shorted (e.g. by drifts of blowing sand) or improperly installed.  While vibration isolation of equipment is straightforward, acoustic isolation is often much more difficult and can ``short circuit'' well isolated systems. We have attempted to place sources of acoustic noise in separate rooms for this reason.

\subsubsection{Humidity.}

Studies during the first two observing runs showed that periods of very low humidity inside of the buildings, associated with sub-freezing weather, are correlated with high glitch rates in DARM~\cite{alog_blips, Cabero_2019}. One possible explanation is that reduced electrical conductivity associated with dry conditions can increase charge buildup and discharge in electronic systems such as piezo drivers.

\section{GW Event Validation}\label{sec:vetting}

In addition to investigating sources of environmental influences, knowledge acquired from environmental studies contributes to the vetting of GW event candidates. Analysis pipelines search the strain data for astrophysical signals. They are categorized into modeled searches for binary mergers that match the data to template waveforms (e.g. GstLAL~\cite{Cannon_2012} and PyCBC~\cite{Usman_2016}) and unmodeled searches that identify excess energy coherent between multiple detectors (e.g. cWB~\cite{Klimenko_2008}, oLIB~\cite{Lynch_2017}, and BW~\cite{Cornish_2015}).

Contamination of the GW data can occur through any of the means discussed in previous sections. Environmental noise has the potential to be correlated between detectors by stemming from a common source, such as through electromagnetic signals from distant sources or glitches in GPS-correlated electronics. The analysis pipelines estimate the false-alarm probabilities for GW events based on the background rate of randomly coincident events in the detector network. They generate background events by time-shifting the data stream of one detector relative to another by time steps much longer than the light travel time between detectors and longer than the duration of GW signals~\cite{Was_2009}. This method does not account for the possibility of transients being correlated between the detectors due to a common environmental source. Environmental noise is also particularly relevant to un-modeled searches. Unlike template-based methods, these searches make minimal assumptions about the signal waveform and rely more heavily on signal correlation between sites.

The first observation of a GW occurred on 14 Sept 2015~\cite{gw150914}. The event, a short-duration binary black hole merger designated GW150914, required a number of follow-up investigations to find potential noise sources around the time of the event~\cite{Detchar_2016}. This included an examination of the status of all PEM sensors and any significant signals they observed for possible contamination of the GW signal~\cite{Schofield_150914}. A few of the PEM sensors were not working, but because of redundancy, coverage was sufficient. 

Comparisons between Q-transform spectrograms~\cite{Chatterji_2004} of all coincident events in environmental sensors to the time-frequency path of the event revealed that no environmental signals had paths similar to the event candidate. Q-transforms produce a quality-factor-optimized logarithmic tiling of the time-frequency space, making them useful for visualizing transients. The signal-to-noise ratios (SNRs) of the matching signals were also compared to that of the event, showing that even if there were overlapping time-frequency paths, none of the environmental signals were large enough to influence the strain data at the SNR level of the event, based on multiplying the environmental signals by their respective sensor coupling functions.

The validation process for novel events such as GW150914 also includes redundant checks for global sources of environmental noise. We use a dedicated cosmic ray detector located below an input test mass at LHO to examine any association of cosmic ray showers to excess noise in DARM. We also check external observatories for coronal mass ejections, solar radio signals, geomagnetic signals, and RF signals in the detection band as well as higher frequencies.

There was specific concern over a co-incident extremely-high current (504\,kA) lightning strike over Burkina Faso, prompting additional studies of the effects of lightning on the interferometer~\cite{Schofield_lightning}. Investigations of similar strikes found no effect on the strain data and investigations of closer strikes confirmed that the magnetometers were much more sensitive to lightning strikes than the interferometer was. In conclusion there was no reason to veto the first detection based on environmental disturbances.

Subsequent detections throughout O1 and and O2 employed a similar procedure; however the development of the method described in Section~\ref{sec:analysis} for producing coupling functions for all sensors expedited the process. This was especially important for examining environmental noise during GW170817, the first long-duration event detected by LIGO~\cite{gw170817, Schofield_170817}. The longer duration of this event (75\,s) unsurprisingly overlapped with many environmental signals. Based on the coupling functions for those sensors, several of these environmental events were loud enough (estimated DARM signals of up to SNR~4) to have contributed to the interferometer readout, but not enough to account for the GW signal. Furthermore, none of them had a time-frequency morphology that correlated with any features in the candidate signal.

In O3, most of the procedure described above has been automated in order to handle the increase in detection rate. When an event is detected by the astrophysical search pipelines, the omega scan tool~\cite{Davis_2021, Chatterji_2004} is used to search for transient noise in all PEM sensors in the time window spanning the event candidate. It does so by producing a Q-transform for each sensor and reporting those in which there is a transient signal with a false-alarm rate below $10^{-3}$\,Hz. The omega scan also reports the frequency and amplitude of the most significant tile for each sensor. The coupling function of each sensor is interpolated at the peak frequency and multiplied by the peak amplitude to estimate the contribution of the environmental transient to DARM. Sensors whose estimated contribution exceed one tenth of the DARM background level are flagged for human input, requiring a comparison of the environmental signal morphology to that of the event candidate. If there is sufficient signal overlap, reviewers may advise that analysts perform some noise removal in the data, such as by gating or filtering out the appropriate time or frequency range, before performing further follow up analyses. The event could be retracted, if gating or filtering out the environmental contribution would reduce the signal-to-noise ratio of the candidate to a level no longer consistent with a GW detection. During the first half of O3, no candidates were retracted on the basis of the environmental coupling check alone. Some human input was still required for all of the 39 events reported in Abbott et al.\ (2020)~\cite{gwtc2}, although little to no signal overlap of environmental transients was seen.

\section{Conclusions and Future Work}\label{sec:conclusion}

Environmental disturbances continue to be a major topic of investigation in the current generation of gravitational wave detectors. With the transition from initial LIGO to aLIGO, the detectors underwent significant changes, many of which affected their sensitivity to environmental noise. The PEM system for monitoring these noise sources also saw modifications, to account for detector upgrades as well as to expand the coverage of the sensors. Over time we have developed new methods for tracking down noise sources as we have described here. We also described the method for quantifying environmental coupling and its limitations.

As O4 approaches, the detectors are undergoing further upgrades to improve their performance and begin the transition towards the ``A+'' phase~\cite{Miller_2015}. These changes will introduce new hardware and infrastructure, such as a new 300m filter cavity to implement frequency-dependent squeezing~\cite{Evans_2013, McCuller_2020}. The PEM system will continue to be expanded in order to monitor new noise sources that may arise with these upgrades. The installation of wall-mounted magnetic field injection coils will be completed at both sites ahead of O4 to provide full coverage of magnetic coupling. Additionally, shaker injections may also be incorporated in the weekly monitoring program to track changes in low-frequency vibrational coupling.

Further automation to the event validation process will be required to reduce the reliance on human input in future observing runs. This could include providing quantitative estimates on the overlap between the time-frequency path of a PEM signal and an event candidate, as well as estimating the DARM contribution at all times and frequencies rather than just at the time and frequency of the peak sensor amplitude. Environmental monitoring will continue to play a crucial role in the event validation as improved sensitivities bring about higher detection rates and the potential for novel sources of gravitational waves.

\section{Acknowledgements}

LIGO was constructed by the California Institute of Technology and Massachusetts Institute of Technology with funding from the National Science Foundation and operates under Cooperative Agreement No.\ PHY-1764464. Advanced LIGO was built under Grant No.\ PHY-0823459. The authors acknowledge support from NSF grants PHY-1607336, PHY-1912604, PHY-1806656, and PHY-1806656. For this paper, we use the data from the Advanced LIGO detectors and we used the LIGO computing clusters to perform the analysis and calculations.

\newpage
\section*{References}

\bibliography{references}{}
\bibliographystyle{iopart-num}

\end{document}